\documentclass[epsf,epsfig,floatfix,showpacs,groupedaddres,superscriptaddress]{iopart}
\usepackage{iopams}
\usepackage{graphicx}
\usepackage{placeins} 
\usepackage[labelformat=simple]{subcaption}

\usepackage[usenames,dvipsnames]{xcolor}
\usepackage[german,czech,english]{babel}
\usepackage[T1]{fontenc}
\usepackage[utf8]{inputenc}
\expandafter\let\csname equation*\endcsname\relax
\expandafter\let\csname endequation*\endcsname\relax
\usepackage{amsmath}  
\usepackage{amssymb}
\usepackage{amsfonts}
\usepackage{verbatim}
\usepackage{simplewick}
\usepackage{eucal}
\usepackage{enumerate}
\usepackage{eucal}
\usepackage{calrsfs}

\usepackage{physics}
\usepackage{hyperref} 
\PassOptionsToPackage{pagebackref=true}{hyperref}
\PassOptionsToPackage{unicode}{hyperref}
\PassOptionsToPackage{naturalnames}{hyperref}
\hypersetup{
     colorlinks=true,
     citecolor=blue,
     urlcolor=orange,
     linkcolor=red
}
\usepackage{url}
\usepackage{pstricks}
\newrgbcolor{dgreen}{0.0 0.4 0.0}
\newrgbcolor{pink}{1.0 0.0 0.4}
\newrgbcolor{violet}{0.6 0.0 0.4}
\newrgbcolor{orange}{1.0 0.6 0.0}
\newrgbcolor{dred}{0.8 0.1 0.0}
\providecommand{\email}[1]{\footnote{#1}}
\renewcommand{\fref}[1]{Fig.~\ref{#1}}
\renewcommand{\eref}[1]{Eq.~(\ref{#1})}
\newcommand{\be}{\begin{equation}}
\newcommand{\ee}{\end{equation}}
\def\etal{{\it et al.}}
\def\h{\hat}
\def\g{\gamma}
\def\sig{\sigma}
\def\eps{\epsilon}

\def\bi{\begin{itemize}}
\def\ei{\end{itemize}}
\def\ua{\uparrow}
\def\da{\downarrow}
\def\bnu{\begin{enumerate}}
\def\enu{\end{enumerate}}
\def\i{\item}
\def\non{\nonumber}
\def\blgn#1\elgn{\begin{align}#1\end{align}}
\providecommand{\y}{^{\dag}}
\providecommand{\py}{^{\phantom \dag}}
\def\Ra{\Rightarrow}
\def\tp{t_+}
\def\tm{t_-}
\def\Ep{E^+}
\def\Em{E^-}
\def\Epm{E^\pm}
\def\Ez{E^0}
\def\epsp{\eps_+}
\def\epsm{\eps_-}
\def\TLS{\text{TLS}}
\def\mix{\text{mix}}
\def\re{\text{Re}}
\def\im{\text{Im}}

\def\int{\text{int}}
\def\h{\hat}
\def\sig{\sigma}
\def\is{{i\sigma}}

\def\eps{\epsilon}
\def\bi{\begin{itemize}}
\def\ei{\end{itemize}}
\def\ua{\uparrow}
\def\da{\downarrow}
\def\bnu{\begin{enumerate}}
\def\enu{\end{enumerate}}
\def\i{\item}
\def\non{\nonumber}

\def\hf{\frac{1}{2}}
\def\PT{\mathcal{PT}}

\def\P{\mathcal{P}}
\def\T{\mathcal{T}}

\def\l{\lambda}
\def\inv{^{-1}}
\def\col{\textcolor{black}}
\def\ccol{\textcolor{black}}

\def\FIGDIR{.}
\begin{document}
\title[A tale of two kinds of exceptional point in a hydrogen molecule]{A tale of two kinds of exceptional point in a hydrogen molecule}

\author{Himadri Barman}\email{hbarhbar@gmail.com}
\address{Department of Physics, Zhejiang University, Hangzhou 310027, China}
\author{Suriyaa Valliapan}\email{suriyaadan@gmail.com }
\address{Department Of Physics, IIT Madras, Adyar, Chennai 600036, India}
%
%
\begin{abstract}
We study the parity and time-reversal ($\PT$) symmetric quantum physics in
a non-Hermitian non-relativistic hydrogen molecule with local (Hubbard type) Coulomb interaction. We consider non-Hermiticity generated from both kinetic and orbital energies of the atoms and encounter the existence of two different types of exceptional points (EPs) in pairs. These two kinds of EP are characteristically different and depend differently on the interaction strength. Our discovery may open the gates of a rich physics emerging out of a simple Hamiltonian resembling a two-site Hubbard model.
\end{abstract}

\maketitle

%
%
\section{Introduction}
In traditional quantum physics courses at the undergraduate level, only linear Hermitian operators are discussed, keeping the conventional wisdom that a quantum observable in a measurement experiment must possess real eigenvalues and the Hermiticity property of the observable ensures that. However, in 1998, Bender and Boettcher~\cite{bender:boettcher:prl98} showed that Hermiticity is not a necessary condition (though sufficient) for an observable (say, Hamiltonian) to have real eigenvalues. If a Hamiltonian preserves the parity ($\P$) and time-reversal ($\T$) symmetry, it still can exhibit real eigenvalues or eigenenergies within a certain parameter regime. Such Hamiltonians are dubbed $\PT$ symmetric Hamiltonians. As just mentioned, beyond one or more particular points in the parameter space, the Hamiltonian starts picking up complex eigenenergies and those special points are labeled as \emph{exceptional points} (EPs). An EP is the degeneracy point where the complex eigenenergies coalesce. However, unlike the Hermitian degeneracy point, the eigenfunctions become identical (up to a phase factor) instead of being orthogonal to each other. 
\col{
The emergence of complex eigenenergies at the EP indicates the onset of an instability, often dubbed the dynamical instability, resulting from a competition between two or more energy scales involved in the system~\cite{bernier:torre:demler:prl14}.} 
EPs have been interesting for the past decades as they have been the points signaling phase transitions ($\PT$ broken). EPs can signal several exotic phenomena such as unidirectional invisibility~\cite{lin:etal:christo:gr:prl11,regensburger:etal:nat12,zhu:etal:ol13,feng:etal:nmat13}, loss-induced transparency~\cite{guo:etal:prl09}, topological mode switching or energy transfer~\cite{liu:etal:pra21,geng:etal:prsa21}, single mode lasing operation~\cite{hodaei:etal:sc14,feng:etal:sc14}, on-chip control of light propagation~\cite{peng:etal:nphys14}, optical sensitivity against external perturbation~\cite{wiersig:prl14,lin:etal:christo:gr:prl11,hodaei:etal:nat17}, and dynamic phase transition in condensed matter systems~\cite{tripathi:galda:barman:vinokur:prb16}.  
\col{
Despite a large volume of work devoted to studying $\PT$ symmetry physics in various optical systems and other diverse areas of physics, limited attention has been offered towards its applications in molecular physics and chemistry~\cite{lefebvre:etal:prl09,wrona:etal:srep20,marie:burton:loos:jpcm21}. 
}
Moreover, many-body interaction or correlation effect in condensed matter systems has become a growing interest~\cite{ashida:furukawa:ueda:ncomm17,laurenco:etal:prb18,yoshida:peters:kawakami:hatsugai:prb19,tripathi:galda:barman:vinokur:prb16,tripathi:vinokur:srep20,pan:wang:cui:chen:pra20} with the advent of ultracold atoms~\cite{cheng:etal:rmp10,takasu:etal:ptep20} where interaction can be precisely tuned by the Feshbach resonance. In this aspect, the Hubbard hydrogen molecule is one of the minimal models that can capture the fermionic correlation effect and represents the dimer limit of the Mott physics. Such systems, when interact with an environment in an open quantum system, can lead to an effective non-Hermitian two-level or two-state system (TLS)~\cite{book:feynman:leighton:sands77:iii}. As we relate our results to a TLS eigen spectrum in the forthcoming discussions, we first provide a quick pedagogical introduction of two kinds of non-Hermitian TLS below.
       
A simple TLS can be defined by the following $2\times 2$ matrix.
\begin{align}
{\bf H}_\TLS=
  \begin{bmatrix}
   \eps_1 & 0\\
   0     & \eps_2 
  \end{bmatrix}\,.
\end{align}
Here the eigenenergies $\eps_1$ and $\eps_2$ denote the two separate quantum states (if $\eps_1\ne \eps_2$) or degenerate quantum states (if $\eps_1=\eps_2$).  Now if there is mixing between the separated states (say, due to photon absorption/emission, a particle from the lower/higher energy level reaches the higher/lower energy level), we get a finite off-diagonal term (say, $t$). Then the Hamiltonian looks like
\begin{align}
{\bf H}_\TLS^\mix=
  \begin{bmatrix}
   \eps_1 & t\\
   t     & \eps_2 
  \end{bmatrix}\,.
\end{align}
The mixing Hamiltonian is also known as the Landau-Zener Hamiltonian in the context of avoided level 
crossing~\cite{rubbmark:etal:pra81,shevchenko:ashhab:nori:pr10}.
If $\eps_1$, $\eps_2$, and $t$ are real, ${\bf H}_\TLS$ and ${\bf H}_\TLS^\mix$ are 
Hermitian as they satisfy the Hermiticity condition $a_{ji}^*=a_{ij}$ where $a_{ij}$ is the matrix element at $i$-th row and $j$-th column. Now if we make the diagonal parts complex: $\eps_1=\eps+i\g$ and $\eps_2=\eps-i\gamma$ (gain term $i\g$ and loss term $-i\g$ added to a degenerate energy level $\eps$),
we have
\begin{align}
{\bf H}_\TLS^1=
  \begin{bmatrix}
   \eps+i\g & t\\
   t     & \eps-i\g 
  \end{bmatrix}\,
=\eps{\bf 1}+i\g\sig^z+t\sig^x\,.
\label{eq:H:TLS:loss:gain:1}
\end{align}
The Hamiltonian ${\bf H}_\TLS^1$ fails to satisfy the Hermiticity condition and hence non-Hermitian. 
%
%
However, we can easily write down the following eigenvalue or characteristic equation. 
\blgn
(E-\eps)^2+\g^2-t^2=0
\elgn
providing the eigenenergies:
\blgn
E_1,E_2=\eps\pm \sqrt{t^2-\g^2}\,.
\elgn 
%
Like in the previous example, non-Hermiticity can also be introduced via asymmetry in the off-diagonal terms in the TLS matrix, for example,
\blgn
{\bf H}_{\text{TLS}}^{2}=
  \begin{bmatrix}
   \eps & t+\l\\
   t-\l     & \eps 
  \end{bmatrix}\,
\elgn
leading to the  characteristic equation:
\blgn
(E-\eps)^2=\l^2-t^2
\elgn
which provides the eigenenergies:
\blgn
E_1,E_2=\eps\pm \sqrt{t^2-\l^2}\,.
\label{eq:Es:TLS:2}
\elgn
Despite ${\bf H}_{\text{TLS}}^{1}$ and ${\bf H}_{\text{TLS}}^{2}$ being non-Hermitian, their characteristic
equations show that eigenenergies can become real within certain non-Hermiticity parameter regimes: $|\g|\le t$ and $|\l|\le t$ respectively while these parameters are real. Both these Hamiltonians preserve the $\PT$ symmetry~\cite{wang:ptrsa13} and beyond the above-mentioned regimes, complex eigenenergies emerge leading to $\PT$ symmetry broken phases. In our paper, we shall address both these scenarios and study the nature of EPs. We dub the first kind of Hamiltonian (${\bf H}_{\text{TLS}}^{1}$) \emph{diagonal or orbital} $\PT$-symmetric and the second kind ((${\bf H}_{\text{TLS}}^{2}$) \emph{off-diagonal or kinetic}  $\PT$-symmetric. We construct both of these scenarios in the context of the hydrogen molecule: our testing model.

Our paper is organized in the following way. We first discuss the non-interacting version of the hydrogen molecule
and how the eigenenergies are obtained after constructing the basis set and the Hamiltonian matrix upon that. 
Then we introduce the asymmetry into the hopping elements keeping the $\PT$-symmetry reserved for the 
Hamiltonian and discuss the behavior of its complex eigenenergies. We then introduce the Hubbard interaction
term to that and discuss the complex eigenenergies. In the next section, we add complex gain and loss terms to the orbital energies (maintaining the $\PT$-symmetry again) and discuss the existence of multiple sets of EPs and their dependence on the interaction strength. Finally, we conclude by providing physical interpretation and possible future applications of our findings.

%
%
\section{Noninteracting hydrogen molecule}
 A hydrogen molecule consists of two hydrogen atoms where each atomic electron participates in covalent bonding with the other one. This scenario (neglecting vibrational modes and other interactions) can be modeled by a two-site electronic problem where electrons can hop from one site to another site (mimicking the orbital overlap)~\cite{book:ashcroft:mermin76:ssp,alvarez:blanco:ejp01}. In the second quantization notation, the Hamiltonian is equivalent to the two-site tight-binding Hamiltonian:
\begin{align}
\h H^0 = \eps\sum_\sig (c\y_{1\sig} c\py_{1\sig} + c\y_{2\sig} c\py_{2\sig}) + t\sum_\sig (c\y_{1\sig} c\py_{2\sig} + c\y_{2\sig} c\py_{1\sig})\,
\label{eq:H0}
\end{align}
where $c\y_{i\sig}$ or $c\py_{i\sig}$ operator creates or annihilates an electron of spin $\sig$ at site $i$ ($i\in 1,2$; $\sig \in \ua,\da$) $\big[ c\y_{i\sig}|0\rangle_i=|\sig\rangle_i$; $c\py_{i\sig}|\sig\rangle_i=|0\rangle_i \big]$, $\eps$ is the atomic energy of a hydrogen atom, $t$ is the amplitude of hopping from site 1 to site 2 or vice versa. 

We get six possible atomic states for the above Hamiltonian which form the basis $\{\ket{i}\}$, $i= 1,2,3,4,5,6$, the nonzero matrix elements of the Hamiltonian are (see ~\ref{app:construct:H0})
\blgn
H^0_{11}&=H^0_{22}=H^0_{33}=H^0_{44}=H^0_{55}=H^0_{66}=2\eps\\
H^0_{23}&=t=H^0_{32}\\
H^0_{24}&=-t=H^0_{42}\\
H^0_{35}&=t=H^0_{53}\\
H^0_{45}&=-t=H^0_{54}
\elgn
where $H_{ij}=\bra{i} \h H \ket{j}$ for a generic Hamiltonian matrix element. Thus the Hamiltonian 
appears in the matrix form: 
\begin{align}
{\bf H^0}=
  \begin{bmatrix}
     2\eps &0     &0      &0     &0     &0 \\  
     0     &2\eps &t      &-t    &0     &0 \\  
     0     &t     &2\eps  &0     &t     &0 \\  
     0     &-t    &0      &2\eps &-t    &0 \\  
     0     &0     &t      &-t     &2\eps &0 \\  
     0     &0     &0      &0     &0     &2\eps
  \end{bmatrix}
  \,.
\end{align}

The above matrix can be divided into three block-diagonal matrices and one can note
they represent three distinguished sectors of total spin $S_z=1,0,-1$ (considering each electron 
is a spin-$\frac{1}{2}$ particle):
\begin{align}
{\bf H^0}
&=
\begin{bmatrix}
  ~\boxed{S_z=1} &  &\\
    &\boxed{S_z=0} &\\
    &  &\boxed{S_z=-1}
\end{bmatrix}
\,.
\end{align}
For $S_z=\pm 1$, the eigenenergies are trivial: $E=2\eps$. 
For $S_z=0$ matrix:
\blgn
\begin{bmatrix}
  2\eps &t      &-t    &0\\  
  t     &2\eps  &0     &t \\  
   -t   &0      &2\eps &-t \\  
  0     &t      &-t     &2\eps\\  
\end{bmatrix}
\,,
\label{eq:Sz:0:matrix}
\elgn
the characteristic equation becomes 
\blgn
\begin{vmatrix}
  2\eps-E &t      &-t    &0\\  
  t     &2\eps-E  &0     &t \\  
   -t   &0      &2\eps-E &-t \\  
  0     &t      &-t     &2\eps-E\\  
\end{vmatrix}
=0\,
\elgn
\blgn
\Ra (2\eps-E)^2[(2\eps-E)^2 - 4t^2]=0
\elgn
solving which we obtain the following eigenenergies:
$2\eps$ (degeneracy = 4 for the full Hamiltonian $H^0$), $2(\eps-t)$, and $2(\eps+t)$.
By setting $\eps$ to 0, we get: $0$, $-2t$, and $2t$ as three distinct eigenenergies.
For positive values of $t$, the states with eigenenergy $\pm 2t$ correspond to 
antibonding (energy  $> \eps$) and bonding states (energy $< \eps$) respectively.

%
%
\section{Non-interacting hydrogen molecule with off-diagonal $\PT$ symmetry}
Open quantum systems or dissipative systems have been studied for a long time 
where non-Hermiticity occurs naturally as a decay term in the Hamiltonian~\cite{frensley:rmp90,dalibard:etal92,hatano:nelson:prl96,fukui:kawakami:prb98,bertlmann:etal:pra06}. 
In our model Hamiltonian $H^0$, we introduce non-Hermiticity through the following dissipative current (asymmetric hopping) term $H^\l$~\cite{cabib:prb75}.
\blgn
\h H^\l = \l  \sum_\sig(c\y_{1\sig} c\py_{2\sig} - c\y_{2\sig} c\py_{1\sig})\,.
\elgn
One can easily check that
$\h H\y_\l=\l  \sum_\sig(c\y_{2\sig} c\py_{1\sig} - c\y_{1\sig} c\py_{2\sig})\ne \h H^\l$.
We rewrite our new Hamiltonian as
\blgn
\h H^1
&=H^0+H^\l\non\\
&= \eps\sum_\sig (c\y_{1\sig} c\py_{1\sig} + c\y_{2\sig} c\py_{2\sig}) + \sum_\sig[t^+ c\y_{1\sig} c\py_{2\sig} + t^{-}c\y_{2\sig} c\py_{1\sig}]\,
\label{eq:H:primed}
\elgn 
where $t^+\equiv t+\l;\quad t^-\equiv t-\l$.

\subsection*{$\PT$ symmetry:} 
Since $\h H$ is already Hermitian and hence also $\PT$ symmetric, to prove that $\h H^1$ 
is $\PT$ symmetric as well, we only need to show that $\h H^\l$ is $\PT$ symmetric. %
$\l$ is equivalent to a hopping amplitude and hence it changes sign under time-reversal:  
\blgn
\T \h H^\l \T\inv = -\l \sum_\sig(c\y_{1\sig} c\py_{2\sig} - c\y_{2\sig} c\py_{1\sig})\,. 
\elgn
Now under parity ($\P$) operation, site 1 and 2 get interchanged and we finally obtain 
\blgn
\P\T \h H^\l\T\inv\P\inv = -\l \sum_\sig(c\y_{2\sig} c\py_{1\sig} - c\y_{1\sig} c\py_{2\sig})  
= \h H^\l\,. 
\elgn 
Hence $\h H^\l$ is invariant under $\PT$ symmetry operation and 
the Hamiltonian in matrix form: 
\begin{align}
{\bf H^1}=
  \begin{bmatrix}
     2\eps &0     &0      &0     &0     &0 \\  
     0     &2\eps &t^-      &-t^-    &0     &0 \\  
     0     &t^+     &2\eps  &0     &t^-     &0 \\  
     0     &-t^+    &0      &2\eps &-t^-    &0 \\  
     0     &0     &t^+      &-t^+     &2\eps &0 \\  
     0     &0     &0      &0     &0     &2\eps
  \end{bmatrix}
  \,.
\label{eq:H1:matrix}
\end{align}
Like in the earlier case, we find this matrix also bears a block-diagonal form where the blocks represent three distinguished sectors of total spin ($S_z$) 1, 0 and -1 respectively.
The characteristic equation of the $S_z=0$ block is 
\blgn
\begin{vmatrix}
  2\eps-E &\tm      &-\tm    &0\\  
  \tp     &2\eps-E  &0     &\tm \\  
   -\tp   &0      &2\eps-E &-\tm \\  
  0     &\tp      &-\tp     &2\eps-E\\  
\end{vmatrix}
= 0\,.
\elgn
\blgn
&\Ra (2\eps-E)^2[(2\eps-E)^2-2\tp\tm]
-2\tm\tp(2\eps-E)^2=0\non\\
&\Ra (2\eps-E)^2[(2\eps-E)^2-4\tp\tm]=0\,.
\label{eq:sing:diss:nonint}
\elgn
%
%
%
\begin{figure}[hp!]
\begin{subfigure}[t]{.5\linewidth}
\centering
\includegraphics[height=5cm,clip]{\FIGDIR/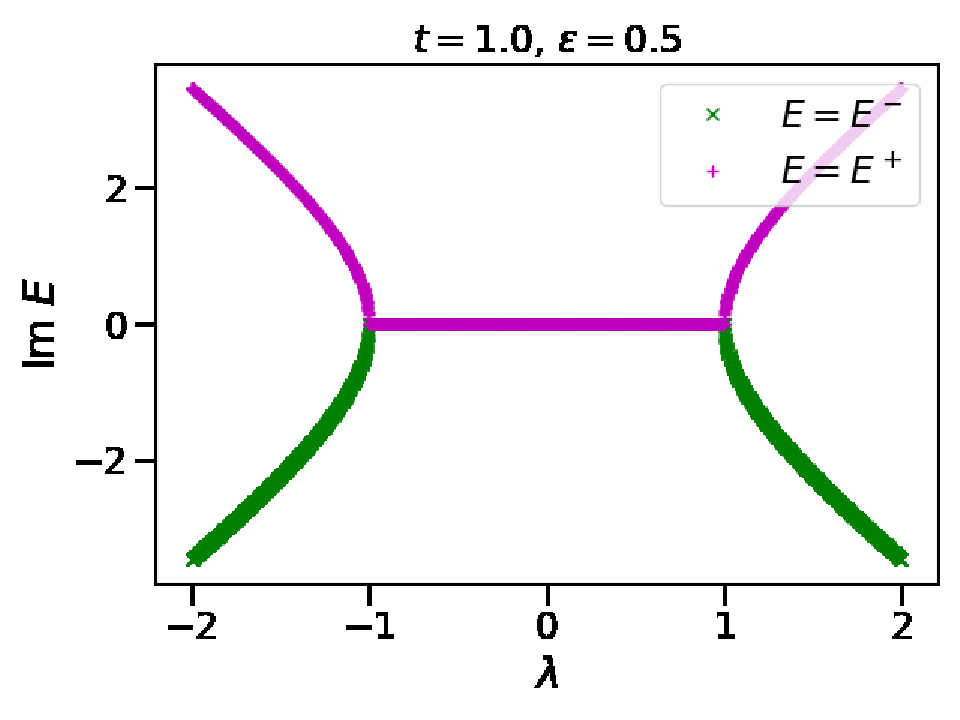}
\caption{}
\label{fig:ImE:vs:lambda:nonint}
\end{subfigure}
\begin{subfigure}[t]{.5\linewidth}
\centering\includegraphics[height=5cm,clip]{\FIGDIR/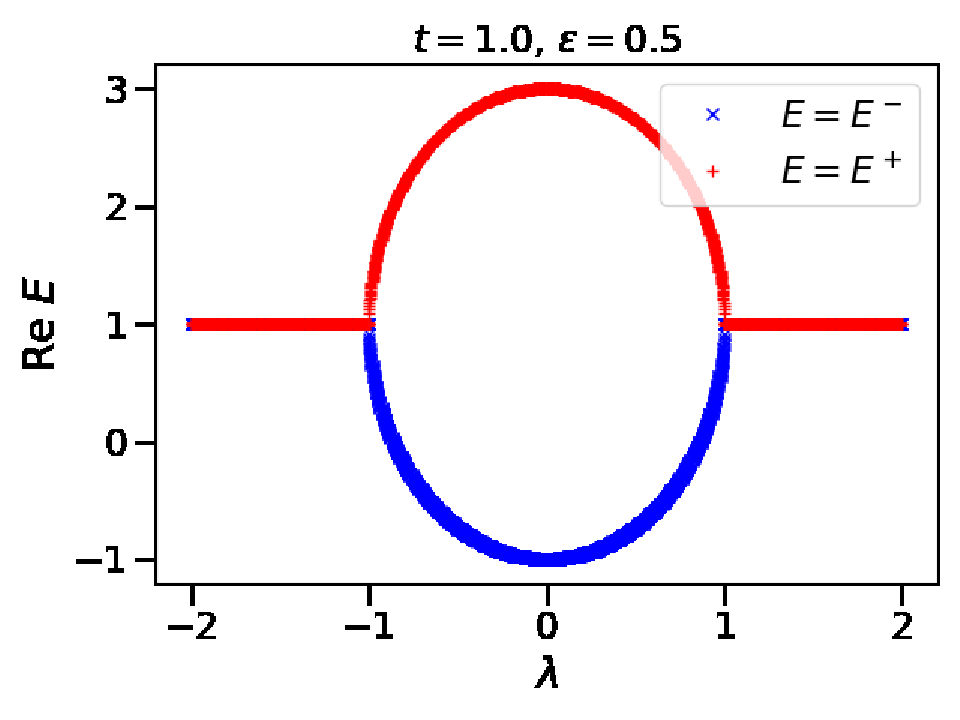}
\caption{}
\label{fig:ReE:vs:lambda:nonint}
\end{subfigure}
\caption{(a) Imaginary and (b) real parts of the two complex eigenenergies of the Hamiltonian $H^1$ plotted as functions of $\l$ for $t=1.0$, $\eps=0.5$.}
\label{fig:E:vs:lambda:nonint}
\end{figure}
%
\FloatBarrier 
Thus the eigenenergies of $\h H^1$ are $2\eps$ (degeneracy 4), $2(\eps\pm \sqrt{t^2-\l^2})$. 
When $|\l|>t$ situation occurs, the last two eigenenergies (we name this pair as $E^\pm$) 
become complex: $E^\pm=2(\eps\pm i\sqrt{\l^2-t^2})$.
Thus symmetrically around $\l=0$, a pair of EPs arise at $\l_e=\pm t$ in the parameter space of $\l$. 
%
%
In \fref{fig:ImE:vs:lambda:nonint}, we plot the real and imaginary parts of $E^\pm$ as functions of $\l$. For our parameter choice $t=1$ and $\eps=0.5$, we find at $|\l|\ge t$, the real parts become zero and the imaginary parts become finite, signifying EPs at $\l_e=\pm t=\pm 1$. The eigenenergies are very similar to that of the typical TLS Hamiltonian in ~\eref{eq:Es:TLS:2} discussed in the Introduction.   
%
%
\col{
The occurrence of these EPs can be physically interpreted as the dynamic instability caused by the interplay between regular and asymmetric hopping (represented by $t$ and $\l$ terms here respectively). The off-diagonal hopping can be considered as the gauge term attached to the hopping resulting in asymmetry in hopping~\cite{hatano:nelson:prl96,fukui:kawakami:prb98}. This asymmetric hopping wins over the regular one once the magnitude of $\l$ becomes greater than $t$ and lets the system become dynamically unstable (entirely dissipative). We see how electronic correlation affects this instability and the EPs in the next section. 
}
%

%
%
\section{Hubbard hydrogen molecule with off-diagonal $\PT$ symmetry}
We turn on the Coulomb interaction between the atoms in the hydrogen molecule and for simplicity, we consider it be the on-site Hubbard interaction ($H^U$) which is routinely used in studies of correlated materials~\cite{book:gebhard10:mott:mit,book:ashcroft:mermin76:ssp}.  The Hubbard interaction term is expressed as
\blgn
H^U \equiv U(\hat{n}_{1\ua}\hat{n}_{1\da}+\hat{n}_{2\ua}\hat{n}_{2\da})
\elgn
where $\h n_{i\sigma}$ is the occupation number operator (${\h n}_\is=c\y_\is c_\is$) and $U$ 
amounts to the Coulomb energy one must pay to bring two electrons of opposite spins together. 
The full interacting Hamiltonian then becomes
\blgn
H^2 = H^0+ H^\l + H^U = H^1 + H^U\,.
\elgn
Since $\h n_{i\sigma}$ is the occupation number operator, we can easily notice
\blgn
H^2\ket{1}&=0\\
H^2\ket{2}&=U\ket{2}\\
H^2\ket{3}&=0\\
H^2\ket{4}&=0\\
H^2\ket{5}&=U\ket{5}\\
H^2\ket{6}&=0\,.
\elgn
Working with the same basis states as before, the total Hamiltonian in matrix form can be written as the sum of the respective matrices for $H^U$ and $H^1$:
\begin{align}
{\bf H^2}
=
  \begin{bmatrix}
     2\eps &0     &0      &0     &0     &0 \\  
     0     &2\eps+U &t^-      &-t^-    &0     &0 \\  
     0     &t^+     &2\eps  &0     &t^-     &0 \\  
     0     &-t^+    &0      &2\eps &-t^-    &0 \\  
     0     &0     &t^+      &-t^+     &2\eps+U &0 \\  
     0     &0     &0      &0     &0     &2\eps
  \end{bmatrix}
  \,.
\label{eq:H2:Sz:0:matrix}
\end{align}
 The characteristic equation for the $S_z=0$ sector of \eref{eq:H2:Sz:0:matrix} is 
\blgn
\begin{vmatrix}
  2\eps+U-E &\tm      &-\tm    &0\\  
  \tp     &2\eps-E  &0     &\tm \\  
   -\tp   &0      &2\eps-E &-\tm \\  
  0     &\tp      &-\tp     &2\eps+U-E\\  
\end{vmatrix}
=0
\non
\elgn
\blgn
&\Ra (2\eps-E)(2\eps+U-E)\non\\
&\qquad\times\bigg[(2\eps-E)(2\eps+U-E)-4\tp\tm\bigg]=0
\label{eq:sing:diss:Hubb:form0}\\
&\Ra (2\eps-E)(2\eps+U-E)\non\\
&\qquad\times\bigg[(2\eps-E+U/2)^2-U^2/4-4\tp\tm\bigg]=0\,.
\label{eq:sing:diss:Hubb}
\elgn
%
%
%
\begin{figure}[htp!]
\begin{subfigure}[t]{.6\linewidth}
\centering\includegraphics[totalheight=6cm,clip]{\FIGDIR/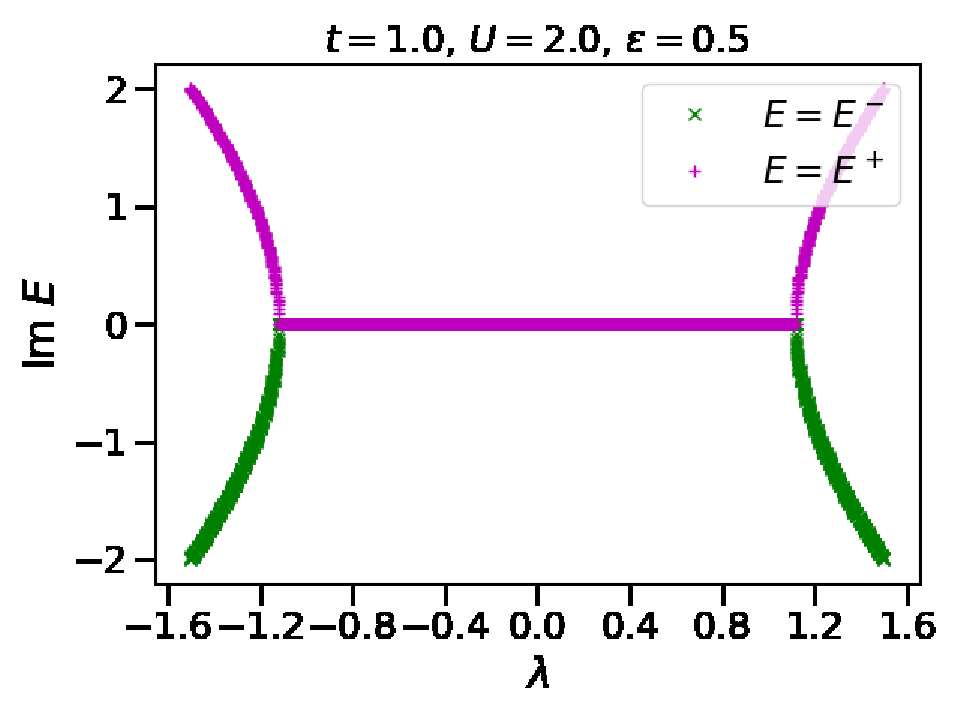}
\caption{}
\label{fig:ImE:vs:lambda:int}
\end{subfigure}
\begin{subfigure}[t]{.6\linewidth}
\centering\includegraphics[totalheight=6cm,clip]{\FIGDIR/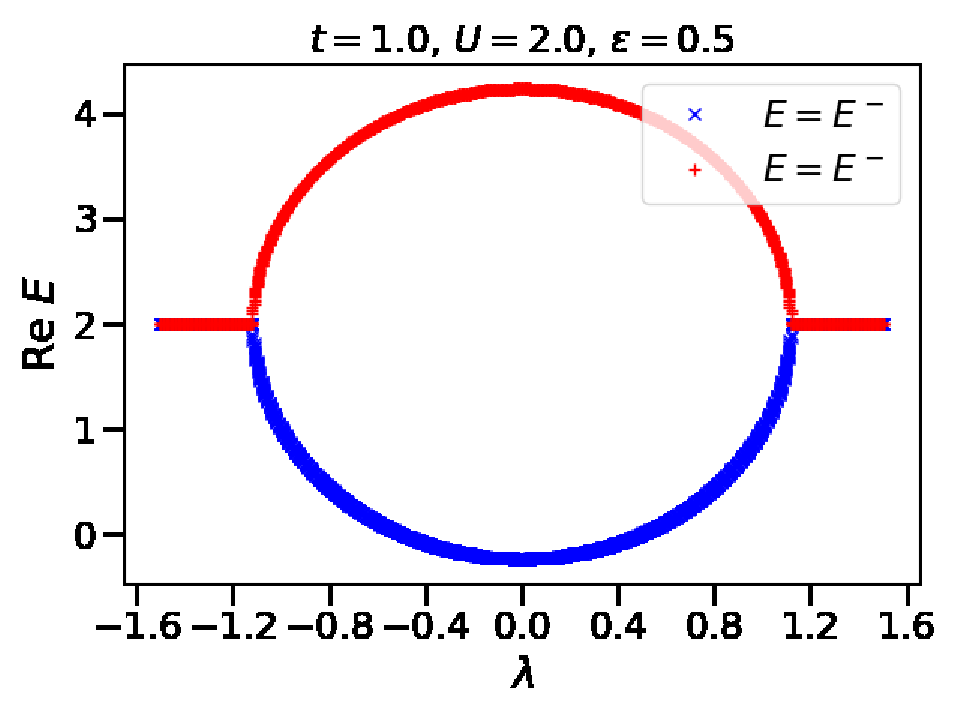}
\caption{}
\label{fig:ReE:vs:lambda:int}
\end{subfigure}
\begin{subfigure}[t]{.5\linewidth}
\centering\includegraphics[totalheight=6cm,clip]{\FIGDIR/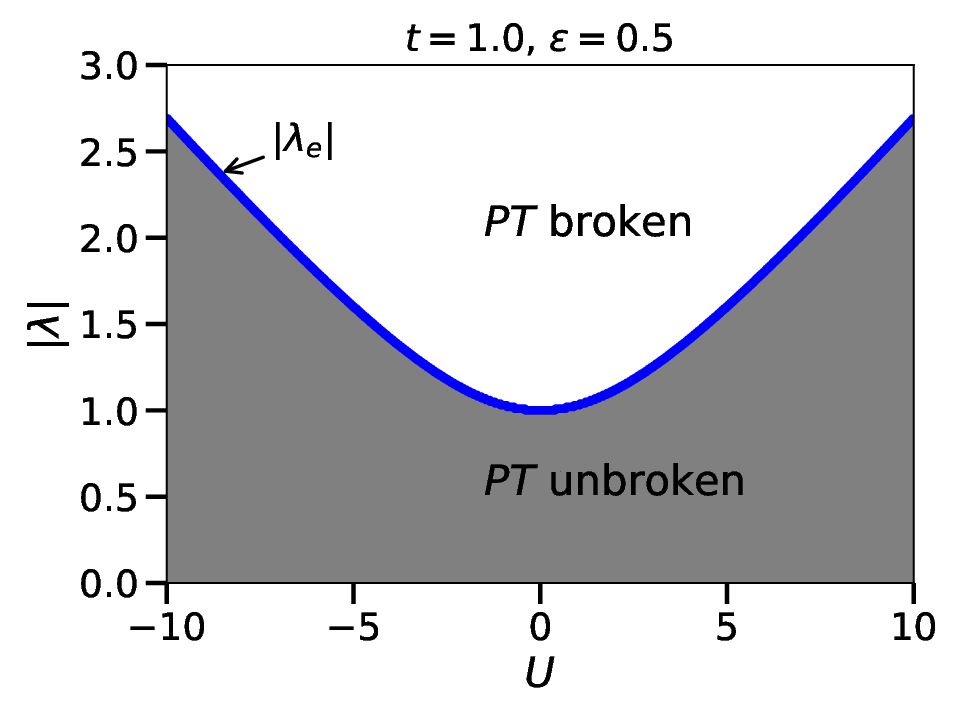}
\caption{}
\label{fig:le:vs:U}
\end{subfigure}
\caption{(a) Imaginary  and (b) real parts of the two complex eigenenergies plotted as functions of $\l$ for $t=1.0$, $\eps=0.5$, and $U=2.0$. (c) The exceptional points positions $|\l_e|$ varying with 
Hubbard interaction strength $U$ marks the boundary between $\PT$ broken and unbroken phases.}
\label{fig:eig:int}
\end{figure}
Thus the eigenenergies of $\h H^2$ are $2\eps$ (degeneracy 3), $2\eps+U$, $\hf(4\eps\pm \sqrt{16\tp\tm + U^2}+U)=\hf(4\eps\pm \sqrt{16(t^2-\l^2) + U^2}+U)$. We can check that by setting $U=0$ in \eref{eq:sing:diss:Hubb}, we get back the non-interacting limit (\eref{eq:sing:diss:nonint}). 
We have complex eigenenergies when the discriminant  (term inside the square root) becomes negative, i.e. when $|\l|> \sqrt{t^2+U^2/16}$. 
Thus presence of interaction shifts the positions of the EPs and we have $\l_e=\pm \sqrt{t^2+U^2/16}$, agreeing with the recently studied Hubbard dimer model with
entirely complex hopping ($t=0$)~\cite{marie:burton:loos:jpcm21}.   
 For our choice of parameters: $t=1$, $U=2$, $\eps=0.5$, we find $\l_e\simeq \pm 1.118$ (see \fref{fig:ImE:vs:lambda:int} and \fref{fig:ReE:vs:lambda:int} for the imaginary and real parts of $E^\pm$).
\fref{fig:le:vs:U} shows $\l_e$ symmetrically shifts from the non-interacting limit ($\l_e(U=0)=1$) as $U$ moves both in positive and negative directions.
%
%
\col{
The presence of finite Coulomb interaction tends to localize electrons, reduces the kinetic energy of the electrons (all kinds of hopping) and prevents the system from encountering the dynamic instability. Thus finite $U$ eventually delays the PT-breaking transition and we notice minimum $|\lambda_e|$ at $U=0$.    
 The parabolic curve for $|\l_e|$ marks the boundary between $\PT$ broken and unbroken phases on the $|\l|-U$ plane. 
}

%
%
\section{Hubbard hydrogen molecule with diagonal $\PT$ symmetry}
We now consider the case when the orbital energies of the hydrogen atoms get tuned to different 
energy levels by addition of complex loss and gain terms. For simplicity, let $\epsp=\eps+i\g$, $\epsm=\eps-i\g$ be the energies, i.e. there are equal amounts of loss and gain terms added to the orbital energies. Hence the orbital part of our Hamiltonian becomes  
\blgn
\h H^\g= \epsp\sum_\sig c\y_{1\sig} c\py_{1\sig} +\epsm\sum_\sig c\y_{2\sig} c\py_{2\sig}\,.
\elgn
Two-level or two-band systems with loss and gain terms have been successfully realized in several photonic and optical setups~\cite{person:rotter:stockmann:barth:prl00,makris:elganainy:christodoulides:musslimani:prl08,guo:etal:prl09,feng:elganainy:ge:nphoton17,li:etal:jpc21}. Considering both diagonal and off-diagonal non-Hermiticity, our most generic $\PT$ symmetric Hamiltonian reads
\blgn
\h H^3 
&= H^\l + H^\g + H^U\non\\
&= \sum_\sig\big[\epsp c\y_{1\sig} c\py_{1\sig} + \epsm  c\y_{2\sig} c\py_{2\sig} 
+  \tp c\y_{1\sig} c\py_{2\sig} + \tm c\y_{2\sig} c\py_{1\sig}\big]\non\\ 
&\quad+ U(\hat{n}_{1\ua}\hat{n}_{1\da}+\hat{n}_{2\ua}\hat{n}_{2\da})\,. 
\label{eq:H3}
\elgn 
\subsection*{$\PT$ symmetry:}
\begin{figure}[!htp!]
\begin{subfigure}[t]{.5\linewidth}
\centering\includegraphics[totalheight=4.3cm,clip]{\FIGDIR/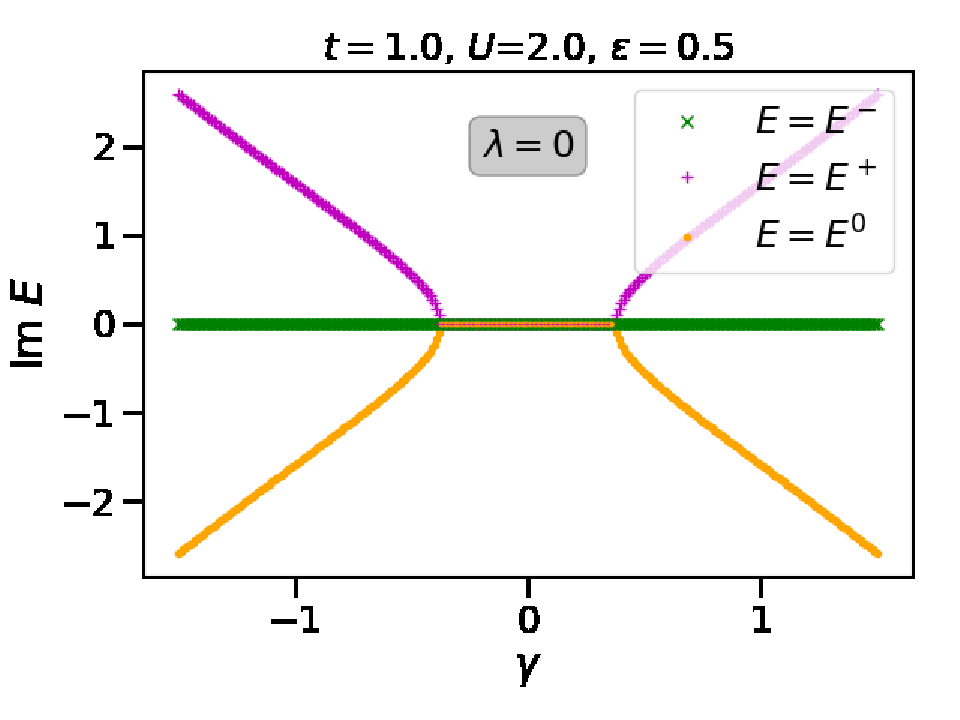}
\caption{}
\label{fig:ImE:vs:gamma:fixed:lambda}
\end{subfigure}
\begin{subfigure}[t]{.5\linewidth}
\centering\includegraphics[totalheight=4.3cm,clip]{\FIGDIR/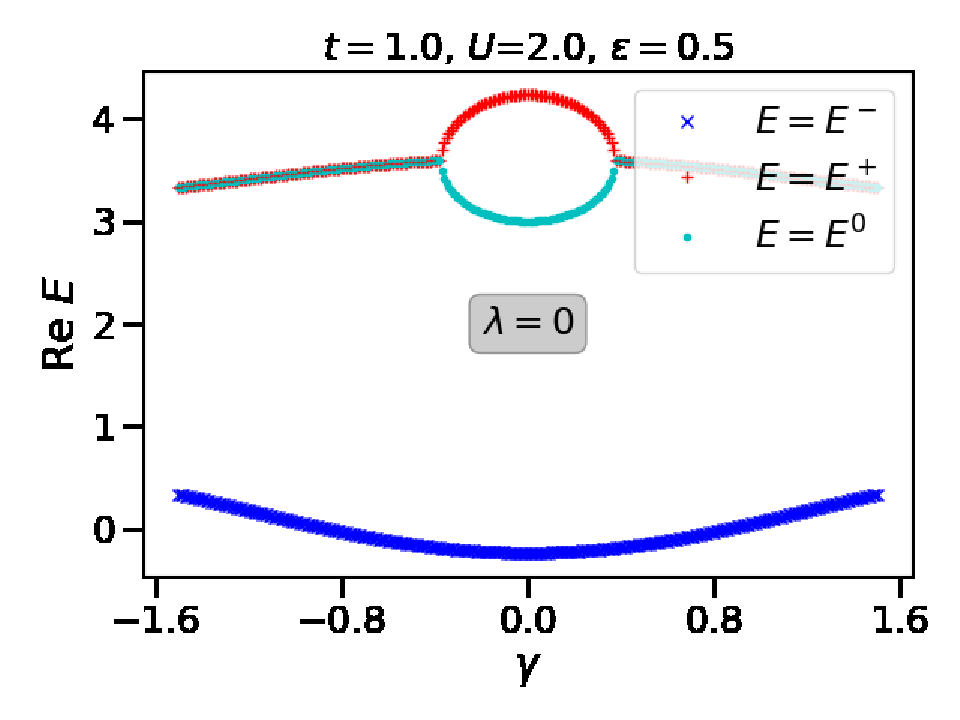}
\caption{}
\label{fig:ReE:vs:gamma:fixed:lambda}
\end{subfigure}
\begin{subfigure}[t]{.5\linewidth}
\centering\includegraphics[totalheight=4.3cm,clip]{\FIGDIR/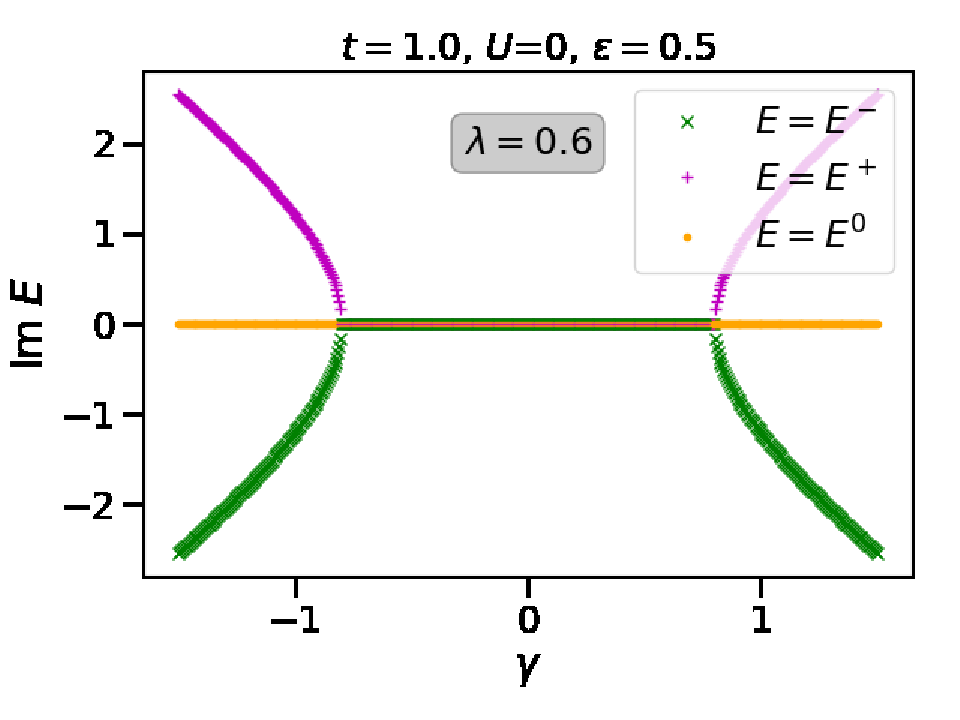}
\caption{}
\label{fig:ImE:vs:gamma:fixed:lambda:U0}
\end{subfigure}
\begin{subfigure}[t]{.5\linewidth}
\centering\includegraphics[totalheight=4.3cm,clip]{\FIGDIR/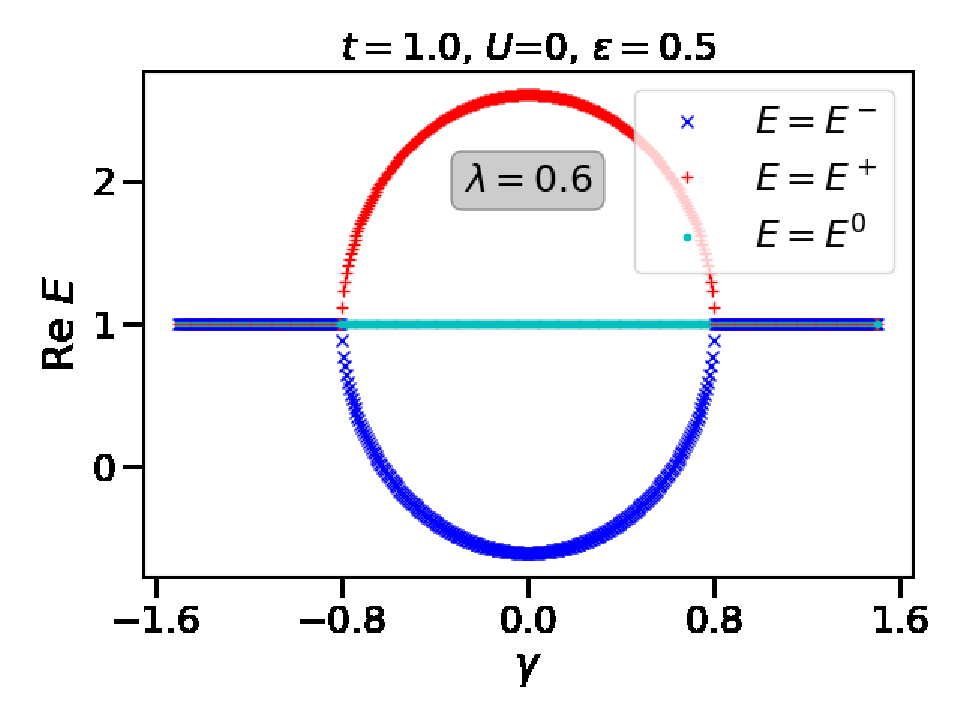}
\caption{}
\label{fig:ReE:vs:gamma:fixed:lambda:U0}
\end{subfigure}
\caption{(a) Imaginary and (b) real parts of the complex eigenenergy bands plotted as functions of  
loss/gain parameter $\gamma$ for $t=1.0$, $\eps=0.5$, and $U=2.0$ at $\l=0$. $\Ez$ and $\Ep$ take part in
forming complex conjugate pair of bands while $\Em$ remains purely real. (c) Imaginary and (d) real parts of the complex eigenenergy bands plotted against $\g$ for the non-interacting case ($U=0$) at $\l=0.6$ while other parameters remain the same. In the non-interacting situation, only $\Epm$ form the complex conjugate band pair while $\Ez=2\eps=1$ remains constant and purely real. Thus the TLS eigenenergy symmetry is recovered.}
\label{fig:E:vs:gamma:fixed:lambda}
\end{figure}

$H^\g$ is $\PT$ symmetric as we can check:
Under $\T$ operation
\blgn
\T H^\g \T\inv= \sum_\sig\big[\epsm c\y_{1\sig} c\py_{1\sig} + \epsp c\y_{2\sig} c\py_{2\sig}\big]   
\elgn
and under $\PT$ operation
\blgn
\P\T H^\g\T\inv \P\inv = \sum_\sig\big[\epsm c\y_{2\sig} c\py_{2\sig} + \epsp c\y_{1\sig} c\py_{1\sig}\big]
= H^\g\,.   
\elgn
Following the same basis formulation, we get the Hamiltonian in matrix form:
\begin{align}
&{\bf H^3}\non\\
&=
  \begin{bmatrix}
     \epsp+\epsm &0     &0      &0     &0     &0 \\  
     0     &2\epsm+U &\tm      &-\tm    &0     &0 \\  
     0     &\tp     &\epsp+\epsm  &0     &\tm     &0 \\  
     0     &-\tp    &0      &\epsp+\epsm &-\tm    &0 \\  
     0     &0     &\tp      &-\tp     &2\epsp+U &0 \\  
     0     &0     &0      &0     &0     &\epsp+\epsm
  \end{bmatrix}
  \,.
\label{eq:H3:Sz:0:matrix}
\end{align}
Again like in the earlier cases, the $S_z=0$ sector of the block-diagonal form
yields the characteristic equation:
\blgn
&(\epsp+\epsm-E)\non\\
&\quad\times\bigg[(2\epsm+U-E)(\epsp+\epsm-E)(2\epsp+U-E)\non\\
&\quad-4\tp\tm(\epsp+\epsm+U-E)\bigg]=0
\label{eq:doub:diss:Hubb:form0}\\
&\Ra (\epsp+\epsm-E)\non\\
&\quad\times\bigg[(2\epsm+U-E)(\epsp+\epsm-E)(2\epsp+U-E)\non\\
&\quad-4\tp\tm(\epsp+\epsm-E)-4\tp\tm U\bigg]=0\,.
\label{eq:doub:diss:Hubb}
\elgn
\eref{eq:doub:diss:Hubb:form0} reproduces \eref{eq:sing:diss:Hubb:form0} once we set $\g=0$ (then we have $\epsp=\epsm=\eps$). The eigenenergies of $\h H^3$ are
$2\eps$ (degeneracy 3), and the three roots of the cubic equation inside the 
bracket of \eref{eq:doub:diss:Hubb}:
\def\esum{S}
\def\ediff{D}
\blgn
&(2\epsm+U-E)(\epsp+\epsm-E)(2\epsp+U-E)\non\\
&\quad-4\tp\tm(\epsp+\epsm-E)-4\tp\tm U = 0\,
\label{eq:cubic:doub:diss:Hubb}
\elgn
which can be simplified as (see ~\ref{app:cubic})
\blgn
X^3-U X^2-K X - L=0
\label{eq:doub:diss:Hubb:cubic:form}
\elgn
with $X\equiv x+U$; $x\equiv \epsp+\epsm-E$; $K\equiv 4(t^2-\g^2-\l^2)$; $L\equiv 4\g^2 U$. 
Thus once we solve for $X$ in \eref{eq:doub:diss:Hubb:cubic:form} by typical Cardano's method~\cite{book:tignol01:galois} or numerically~\cite{book:press:etal02:nrecipe:inC}, we
expect to have at least one real root all the time, the other two roots become complex
conjugates of each other (since the coefficients of $X$ are real) beyond a certain parameter space. This pair of complex conjugate roots give rise to EPs at the parameter space when the complex roots just become real.
Since we introduce two kinds of non-Hermiticity via the orbital energy and the hopping 
terms, it may be natural to expect observing additional EPs. These EPs are different from higher order EPs~\cite{hodaei:etal:nat17}, since we are focusing always
on the pair of energy levels that become complex in certain parameter regimes, while the other level remains real.
%
%
\ccol{However, unlike the $\gamma=0$ or $U=0$ case discussed in previous sections, this time, three of the eigenenergies or roots of \eref{eq:doub:diss:Hubb:cubic:form} participate in generating different EPs at different places. We label the three eigenenergy levels as $\Em$, $\Ez$, and $\Ep$ and plot them as continuous functions of the parameter of interest (see ~\ref{app:marking:convention} for the convention of marking them). When $U=0$ or $\g=0$, $L$ vanishes
in \eref{eq:doub:diss:Hubb:cubic:form} and $X=x+U=0$ becomes a trivial solution of it.  
$E^0=2\eps+U$ (e.g. see \fref{fig:ReE:vs:gamma:fixed:lambda:U0}) is the corresponding eigenenergy for that solution and it remains constant for any values of the non-Hermiticity parameters (independent of $\l$ and $\g$).
} 
%
%
%
\begin{figure}[htp!]
\begin{subfigure}[t]{.5\linewidth}
\centering
\includegraphics[totalheight=5cm,clip]{\FIGDIR/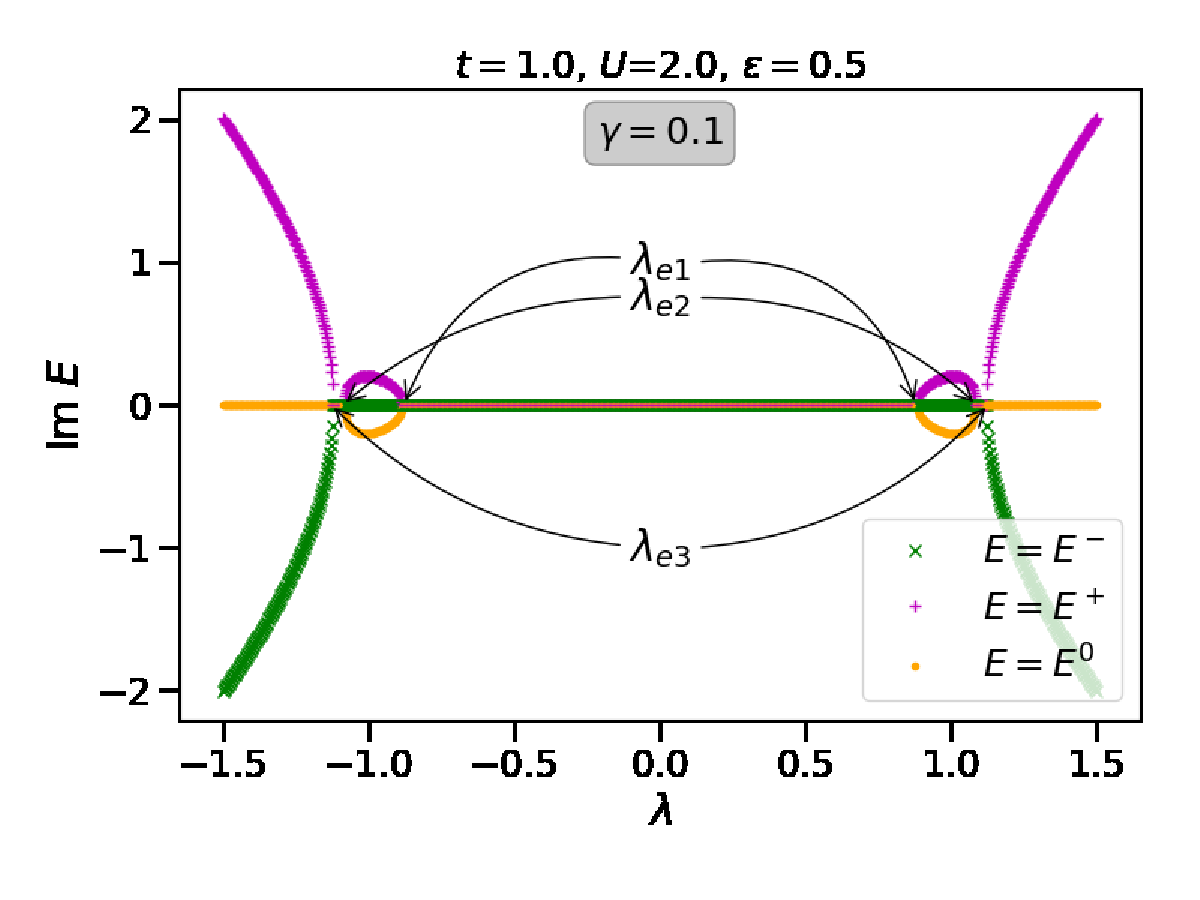}
\caption{}
\label{fig:ImE:vs:lambda:fixed:gamma}
\end{subfigure}
\begin{subfigure}[t]{.5\linewidth}
\centering
\includegraphics[totalheight=5cm,clip]{\FIGDIR/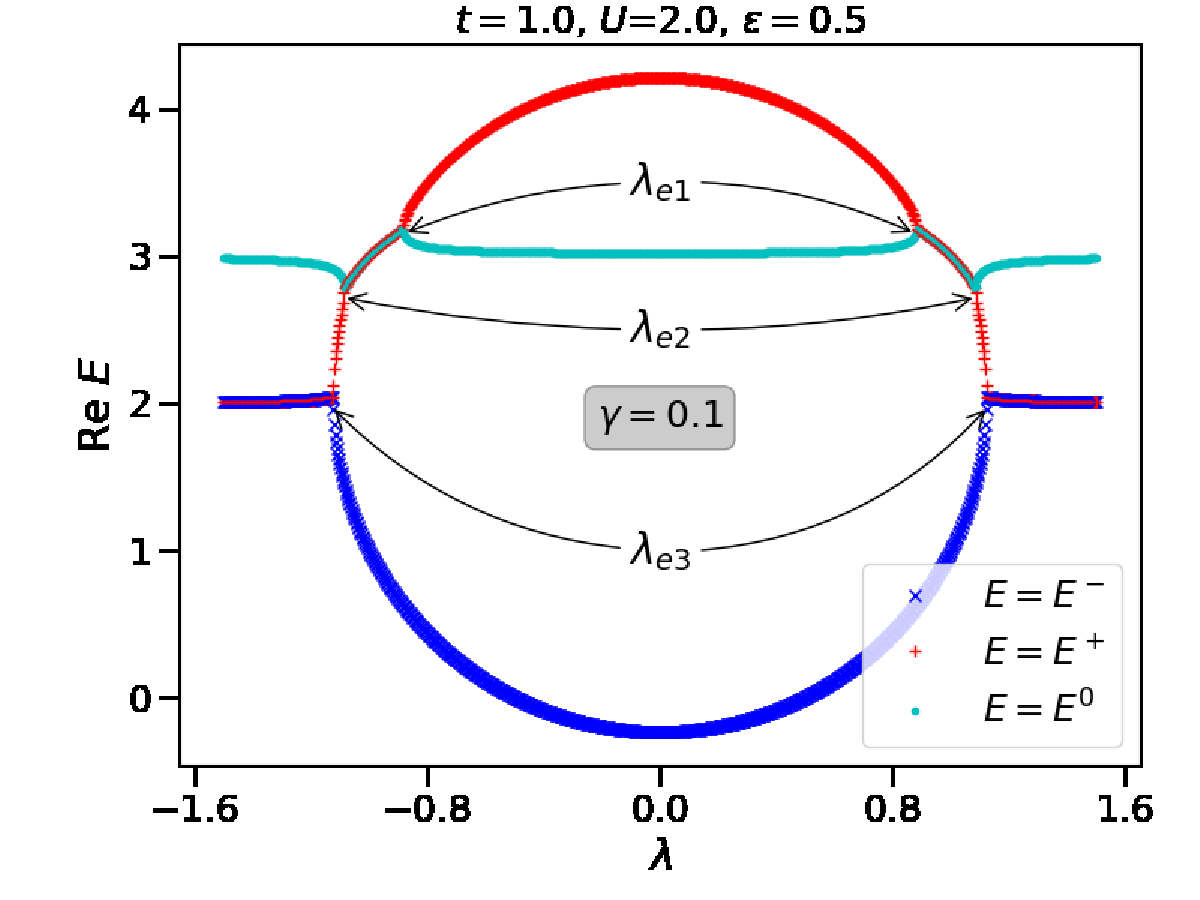}
\caption{}
\label{fig:ReE:vs:lambda:fixed:gamma}
\end{subfigure}
\caption{(a) Imaginary and (b) real parts of the complex eigenenergy bands plotted as functions of  
dissipative parameter $\l$ for $t=1.0$, $\eps=0.5$, and $U=2.0$ at $\gamma=0.1$. The three eigenenergy bands $\Ep$, $\Em$, and $\Ez$ participate in forming complex conujugate pairs on different occasions and create EPs $\l_{e1}$, $\l_{e2}$, and $\l_{e3}$ on both positive and negative sides of $\l=0$.}
\label{fig:E:vs:lambda:fixed:gamma}
\end{figure}
\FloatBarrier
%
\ccol{
We first plot the eigenenergies against the off-diagonal parameter $\l$ keeping the diagonal parameter $\g$ 
fixed. We notice, as we shift $\l$ from zero, imaginary parts of $\Ep$ and $\Ez$  
start becoming finite beyond a point $\l_{e1}$ (corresponding real parts collapse on each other), 
then again disappear at $\l_{e2}$. As the $\l$-point moves further, imaginary parts of $\Em$ and $\Ep$ become
finite above $\l_{e3}$ while their real parts collapse like in the non-interacting case 
 (see \fref{fig:E:vs:lambda:fixed:gamma}). $\l_{e1}$, $\l_{e2}$, and $\l_{e3}$: all these are EPs as they are degenerate onset points of imaginary eigenenergies and like in the previous cases, they appear symmetrically both on the positive and negative sides around $\l=0$. 
Though the presence of additional EPs can be anticipated due to double non-Hermitian terms in the Hamiltonian and cubic nature of the characteristic equation (\eref{eq:doub:diss:Hubb:cubic:form}), the characteristics of all of them are not alike.
Unlike the previous cases, the presence of additional EPs breaks the mirror or reflection symmetry between $\re\,E^\pm$, i.e. the energy levels are no longer equally distributed around the EPs and $E^0$ no longer
remains constant and purely real (see \fref{fig:ReE:vs:lambda:fixed:gamma}).
The different EPs arise from the cubic polynomial nature of the characteristic equation.
In both cubic and quadratic equations, real parts remain the same when two of the solutions are complex conjugate of each other ($\PT$ broken). When all roots are real, in the quadratic case, the real parts have to be always symmetric around the EP value (\emph{Cf.} $x=-b/(2a)$ for the quadratic equation $ax^2 + bx + c = 0$). On the other hand, in the cubic case, such symmetry cannot be guaranteed.
}
%
%
%
%
%
%
%
%

%
\ccol{ 
These additional EPs are different because the eigenenergies are generated from complex conjugate pairs of the roots of a cubic equation, where the discriminant depends on an additional coefficient compared to the quadratic equation's case.
The asymmetry in the real parts of $E^\pm$ gets reversed once we change the sign of $U$. 
The asymmetry becomes more evident when we plot the eigenenergies against $\gamma$ for fixed
$\l$ or even when $H^\l$ is turned off (see \fref{fig:ReE:vs:gamma:fixed:lambda}). \fref{fig:ImE:vs:gamma:fixed:lambda} and \fref{fig:ReE:vs:gamma:fixed:lambda} show that $\Ep$ and $\Ez$ energy bands are solely responsible in creating TLS-type EPs on the $\g$-axis keeping $\Em$ purely real for the entire parameter regime. Again, once we set $U$ negative, the asymmetry is reversed and this time, $\Em$ and $\Ez$ bands create the EPs (figures not shown).
However, when we set $U=0$,
we get back the mirror symmetric pair of $\re\,\Epm$ just like our familiar TLS (see \fref{fig:ReE:vs:gamma:fixed:lambda:U0}). This can be easily understood by noticing that \eref{eq:doub:diss:Hubb:cubic:form} reduces to effectively quadratic equation $x^2-4(t^2-\g^2-\l^2)=0$ (for $t^2\ne \g^2+\l^2$) which produces typical square root EPs at $\g_e=\pm 2\sqrt{t^2-\l^2}$ and in $\l_e=\pm 2\sqrt{t^2-\g^2}$ in $\g$ and $\l$ parameter spaces respectively, similar to the form $\l_e$ has for $H^1$ and $H^2$. 
}
%
%
%
%
%
%
%
\def\myht{7cm}
\begin{figure}[htp!]
\begin{subfigure}[t]{.7\linewidth}
\centering
\hspace*{-.5cm}\includegraphics[height=\myht,clip]{\FIGDIR/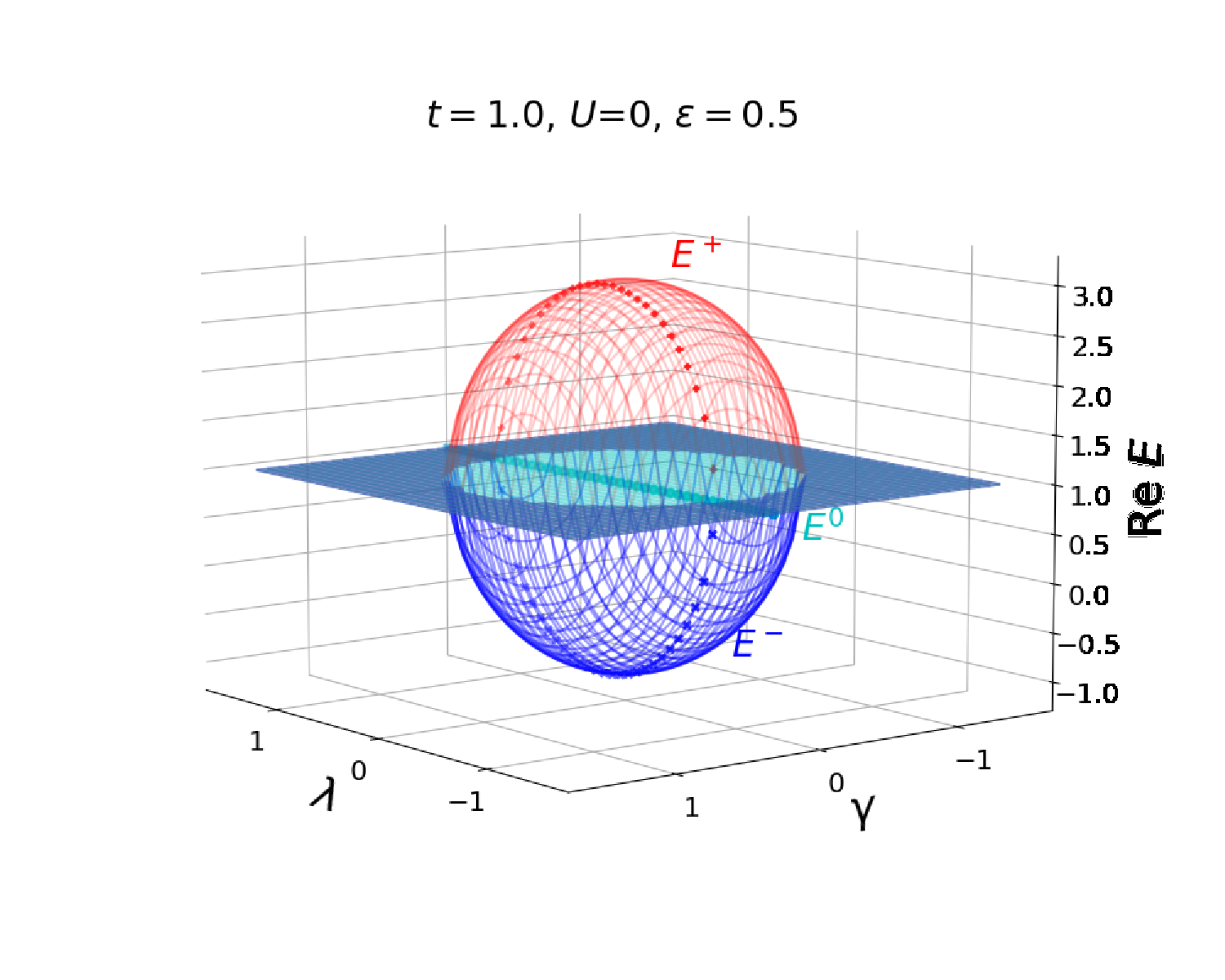}
\caption{}
\label{fig:3D:ReE:U0}
\end{subfigure}
\begin{subfigure}[t]{.5\linewidth}
\centering
\hspace*{-.5cm}\includegraphics[height=\myht,clip]{\FIGDIR/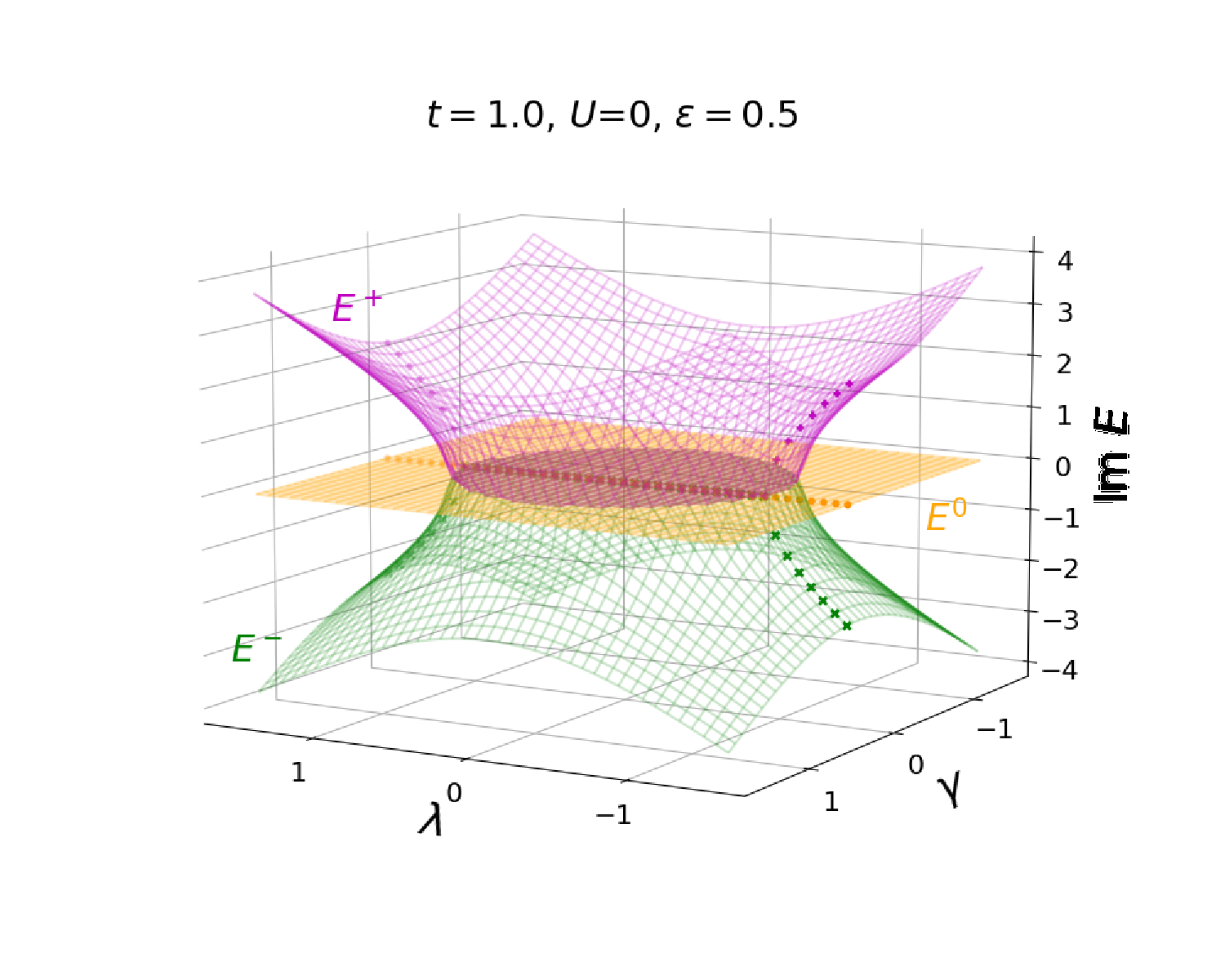}
\caption{}
\label{fig:3D:ImE:U0}
\end{subfigure}
\begin{subfigure}[t]{.7\linewidth}
\centering
\includegraphics[height=\myht,clip]{\FIGDIR/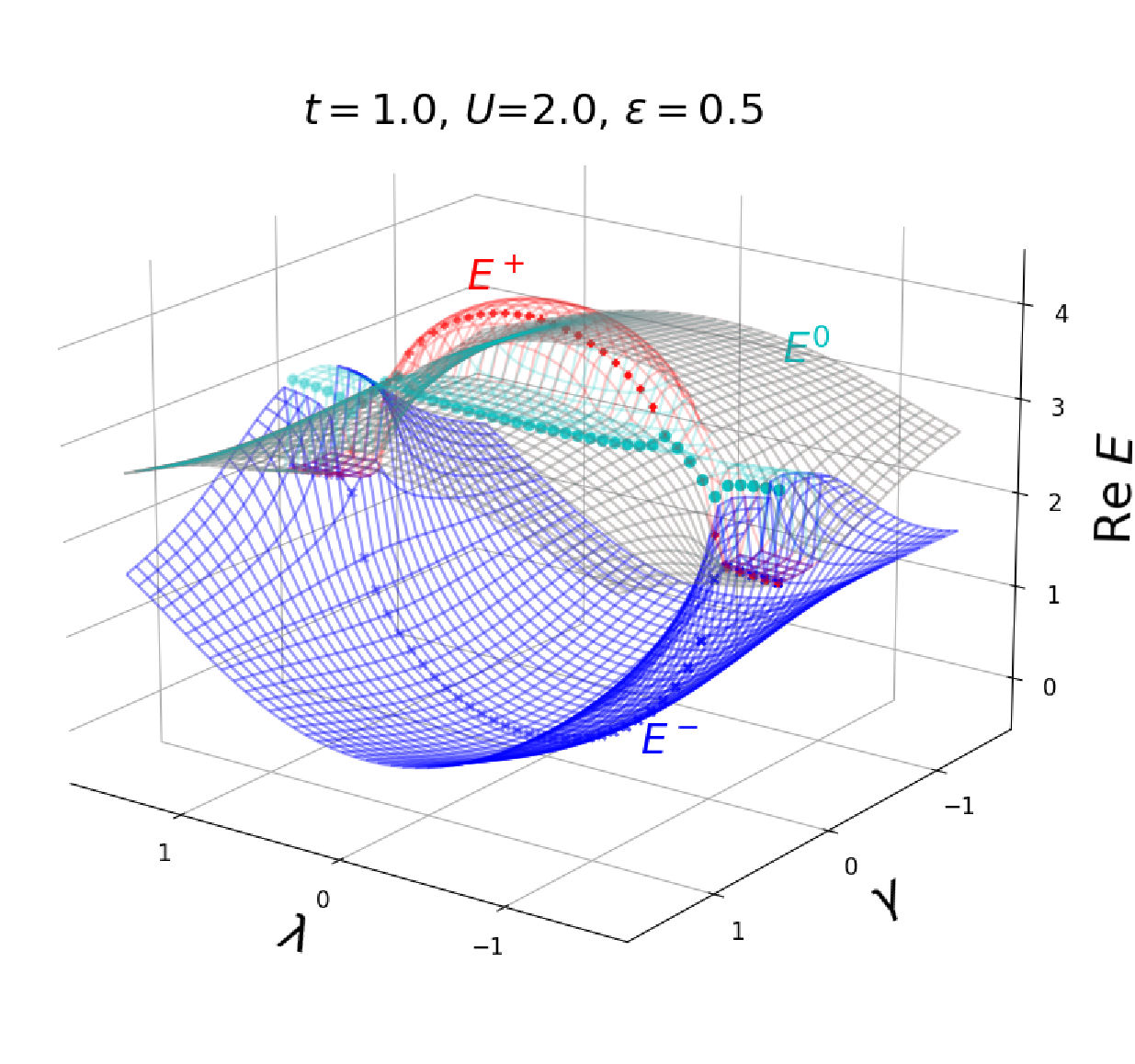}
\caption{}
\label{fig:3D:ReE:U2}
\end{subfigure}
\begin{subfigure}[t]{.5\linewidth}
\centering
\hspace*{.1cm}\includegraphics[height=\myht,clip]{\FIGDIR/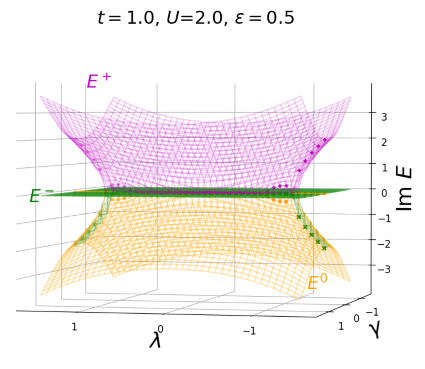}
\caption{}
\label{fig:3D:ImE:U2}
\end{subfigure}
\caption{(a) Real and (b) imaginary parts of eigenenergies plotted over $\l$-$\g$ plane for the hydrogen molecule ($U=0$). Real and imaginary parts of $E^+$ and $E^-$ possess reflection symmetries from the plane $\re\,\Ez=1$ and $\im\,\Ez=0$ respectively.
The real parts are fully rotationally symmetric while the imaginary parts are 4-fold rotationally symmetric on $\l$-$\g$ plane. 
As soon as we turn on the Hubbard interaction, both the rotational and reflection symmetries get broken. (c) and (d) depict the surface plots for $U=2$. Point plots for $\g=0.1$ are provided to guide the eyes. 
}
\label{fig:3D:E:vs:lambda:gamma}
\end{figure}

\ccol{
The scenario becomes much clear when we plot the energy surfaces over $\l$-$\g$ plane for a fixed value of $U$. \fref{fig:3D:ReE:U0} and \fref{fig:3D:ImE:U0} plot real and imaginary parts of the eigenspectra for the non-interacting case ($U=0$). In both plots, we notice that the spectra of $E^+$ and $E^-$ are equally separated from a middle plane ($\re\,\Ez=2\eps=1$ for the real parts and $\im\,\Ez=0$ for the imaginary parts). Clearly, this constitutes a circle of EPs or an \emph{exceptional ring}, centered at ($(\l,\g)=(0,0)$), as a signature of two-dimensional symmetry dependence of the eigenenergies. Thus in absence of the interaction $U$, both non-Hermiticity parameters act on an equal footing with the Hamiltonian despite their dynamics being different (off-diagonal and diagonal). This raises the question of whether our Hamiltonian can have a mapping to another Hamiltonian where the diagonal and off-diagonal parameters can be interchanged without losing the essential physics. We hope the answer to this fundamental question will be found soon by the large community already involved in the diverse research area of non-Hermitian physics. Similar exceptional rings have already been observed in Dirac and Weyl semimetals~\cite{zhen:etal:nat15,yoshida:peters:kawakami:hatsugai:prb19,liu:he:yang:nori:prl21}, however, the non-Hermitian Hamiltonians there do not necessarily require to be $\PT$-symmetric. Now as we turn on $U$, the mirror symmetry between $\Ep$ and $\Em$ gets disrupted and secondary EPs arise along the $\l$-line. The finite $U$ surface plots (\fref{fig:3D:ReE:U2} and \fref{fig:3D:ImE:U2}) show that $\re,\im\,E^0$ surfaces, which no longer remain flat like in the non-interacting case, play a significant role in determining the fates of the EPs discussed already. 
} 

\begin{figure}[htp!]
\begin{subfigure}[t]{.6\linewidth}
\includegraphics[height=6cm,clip]{\FIGDIR/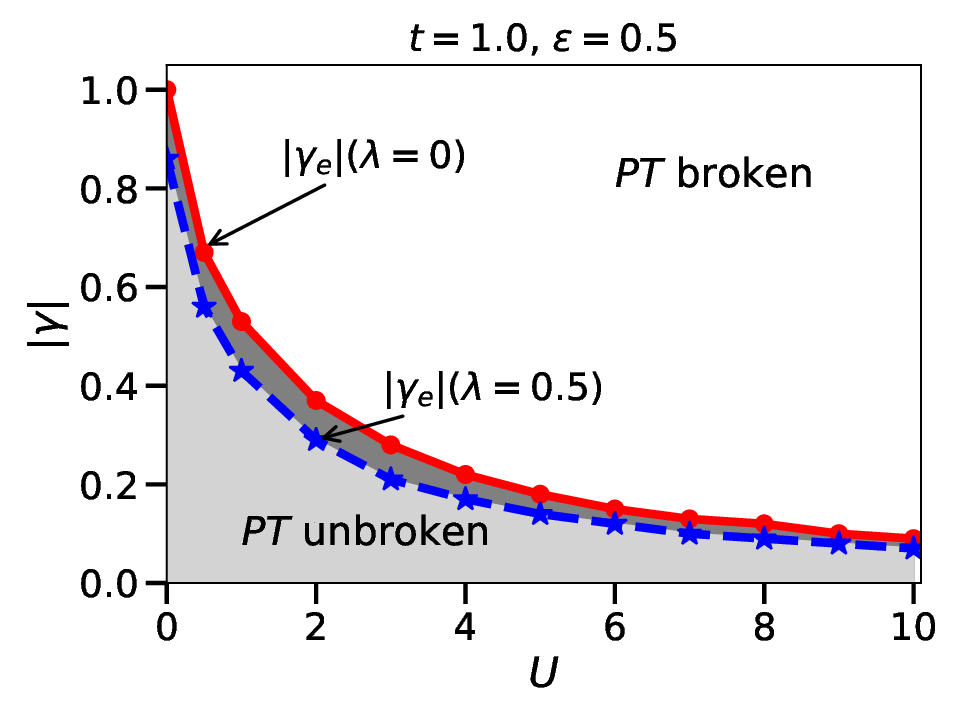}
\caption{}
\label{fig:gamma_e:vs_U}
\end{subfigure}
\begin{subfigure}[t]{.5\linewidth}
\includegraphics[height=7cm,clip]{\FIGDIR/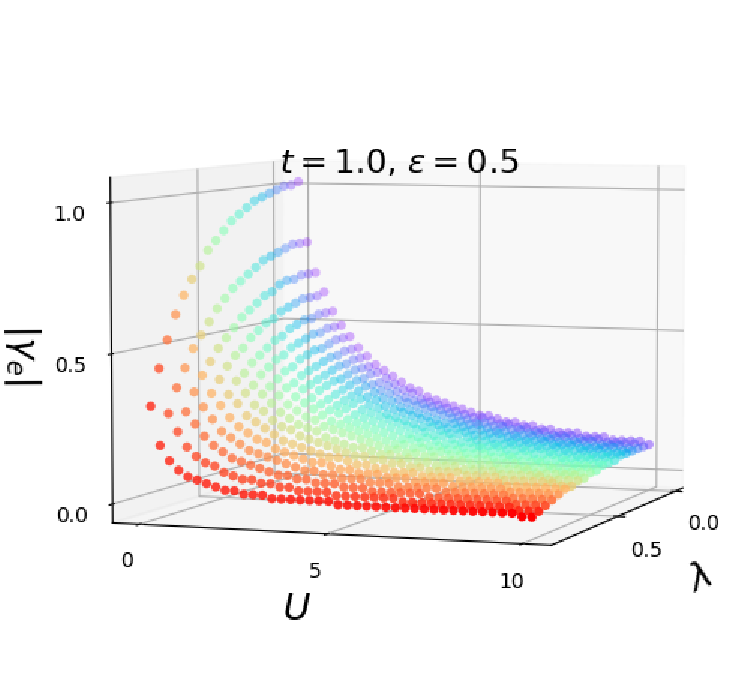}
\caption{}
\label{fig:3Dscat:gammae:vs:l:U}
\end{subfigure}
\caption{(a) $\PT$ broken and unbroken phases on $\g$-$U$ plane for $t=1$, $\eps=0.5$.
The upper and lower curves show the phase boundaries 
(trajectory of $|\g_e|$) for zero and finite ($\lambda=0$) off-diagonal non-Hermiticity parameters. 
(b) 3D scatter plot of $|\g_e|$  over $\lambda$-$U$ plane.}
\end{figure}

\col{
The diagonal non-Hermiticity driven $\PT$ broken and unbroken phase diagrams are shown in \fref{fig:gamma_e:vs_U}. For no other non-Hermiticity parameter ($\l=0$), the phase boundary hits the value of hopping amplitude  ($t=1$ in our case) in the non-interacting limit ($U=0$). This agrees with the result recently obtained by Pan \etal~\cite{pan:wang:cui:chen:pra20}. However, as soon as the off-diagonal non-Hermiticity parameter is turned on (e.g. $\l=0.5$ case shown \fref{fig:gamma_e:vs_U}), the boundary diminishes implying $PT$-symmetry breaking at lower values of $\g$.
\fref{fig:3Dscat:gammae:vs:l:U} plots the trajectory the $|\g_e|$ (EPs on the $\g$ axis) on the $U$-$\l$ plane reflecting the fact that both $U$ and $\l$ diminish its value and hence the $\PT$-broken phase transition gets delayed. This signifies two kinds of competition the $\g$-parameter encounters: (i) off-diagonal non-Hermiticity ($\l$) acting against the diagonal non-Hermiticity and (ii) Hermitian localization parameter ($U$) acting against the destabilizing loss-gain parameter.   
} 
%

%
%
%
\subsection*{\col{Two kinds of exceptional points -- interaction-generated and self-generated:}}
As we notice that the presence of three sets of EPs and interaction plays a role in creating an asymmetry in the real eigenvalues, we decide to plot their positions $\l_{e1}$, $\l_{e2}$, and $\l_{e3}$ against the interaction strength. \fref{fig:lambdae1:vs:U} shows that  $\l_{e1}$ always exists (even when $U=0$) and it decreases as $U$ is increased. On the other hand, \fref{fig:lambdae2:vs:U} and  \fref{fig:lambdae3:vs:U} clearly show that both  $\l_{e2}$ and $\l_{e3}$ arise only at a finite value of $U$ and depending on the value of loss-gain parameter $\g$, it monotonically increases with $U$. $\l_{e3}$'s positions do not vary as significantly as $\l_{e2}$'s do for different $\g$ values (e.g. $\g=0.1$ and $\g=0.2$ shown in the figures). In the non-interacting case, the loop structures in $\im\,E^\pm$ (hence $\l_{e2}$ and $\l_{e3}$) disappear and we only obtain $\l_{e1}$.  \fref{fig:lambdae1:vs:gamma:U}, \fref{fig:lambdae2:vs:gamma:U}, and \fref{fig:lambdae3:vs:gamma:U} plots the EPs against both $\gamma$ and $U$ in three dimensions. The plots show that if we keep the non-Hermiticity parameters fixed, then by tuning the Hubbard interaction $U$, we may encounter EPs as well. Thus the EPs are bounded by the threshold values Hubbard interaction as well~\cite{pan:wang:cui:chen:pra20}. 

\col{The above results allow us to categorize two distinguishable kinds of EPs: (A) \emph{interaction-generated} ($\l_{e2}$ and $\l_{e3}$) and (B) \emph{self-generated}. The interaction-generated EPs are different from traditional EPs often discussed in the literature and deserve special attention and further theoretical and experimental research.
However, crudely we may mention that the EPs are originally the same in nature (as already seen in $U=0$ case) for both non-Hermiticity parameters. It is the presence of interaction that makes the difference between the two. In the gain/loss case, the interaction term appears on the diagonal along with the non-Hermiticity parameter $\gamma$. Hence it is expected to see one kind of EPs depending strongly on $U$ (with the interplay of $\g$ and $U$) while the other one does not (either of $\g=0$ and $U=0$ leads to the same kind of TLS-like eigenenergies). It would be interesting to see the effect of off-diagonal kind of interaction (e.g. nonlocal or extended Hubbard Hamiltonian)~\cite{strack:vollhardt:prl93,loon:etal:prb16} which could be our future work and beyond the scope of this paper.}

%
\def\myht{4cm}
\begin{figure}
	\centering
	\begin{subfigure}[t]{.5\linewidth}
		\centering
		\includegraphics[height=\myht]{\FIGDIR/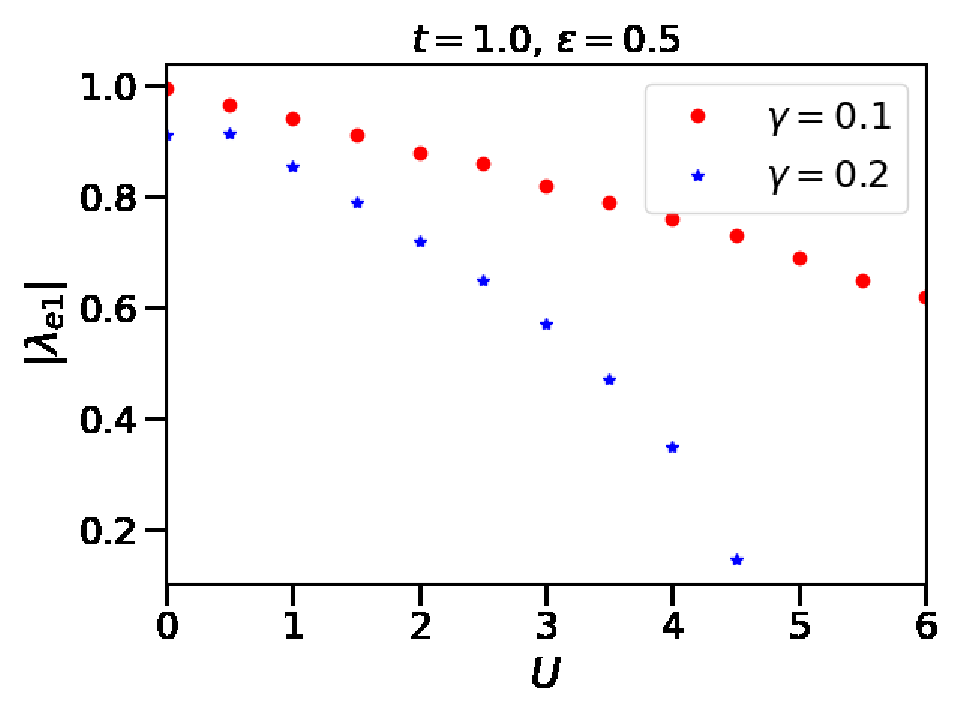}
		 \caption{}
		 \label{fig:lambdae1:vs:U}
	\end{subfigure}%
	\begin{subfigure}[t]{.5\linewidth}
		\centering
                 \includegraphics[height=\myht]{\FIGDIR/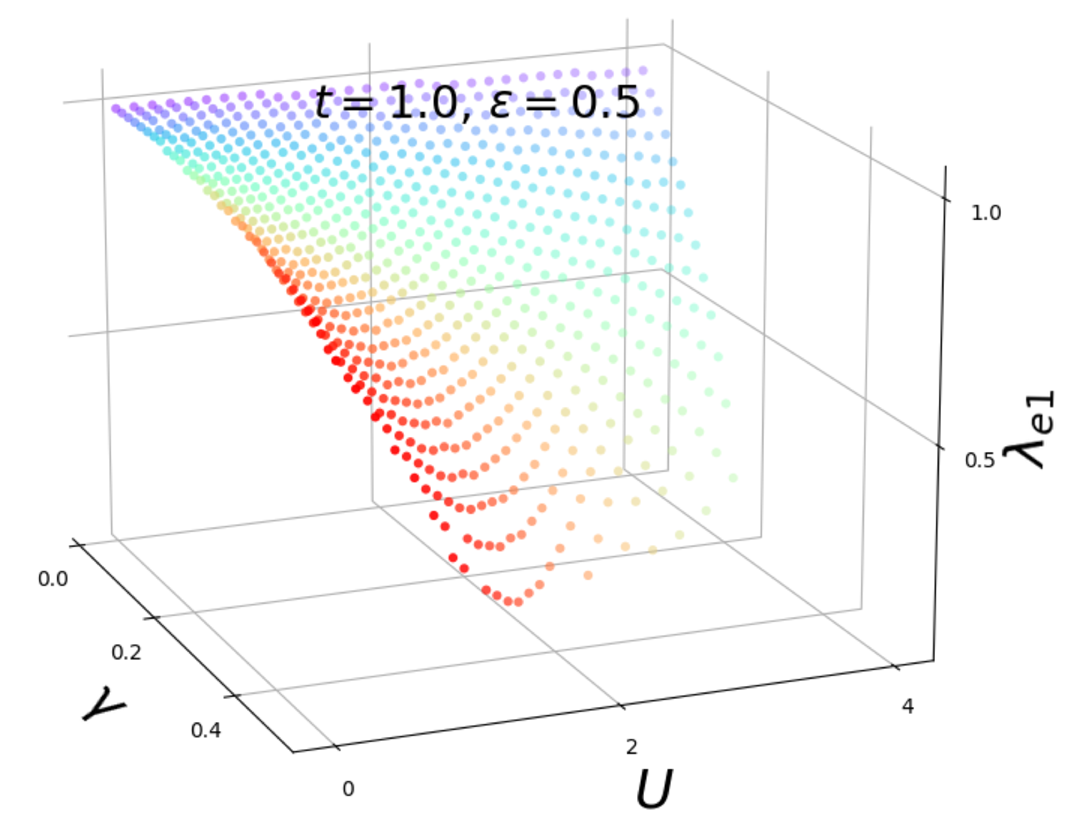}%
		 \caption{}
		 \label{fig:lambdae1:vs:gamma:U}
	\end{subfigure}\\%
        \begin{subfigure}[t]{.5\linewidth}
		\centering
		\includegraphics[height=\myht]{\FIGDIR/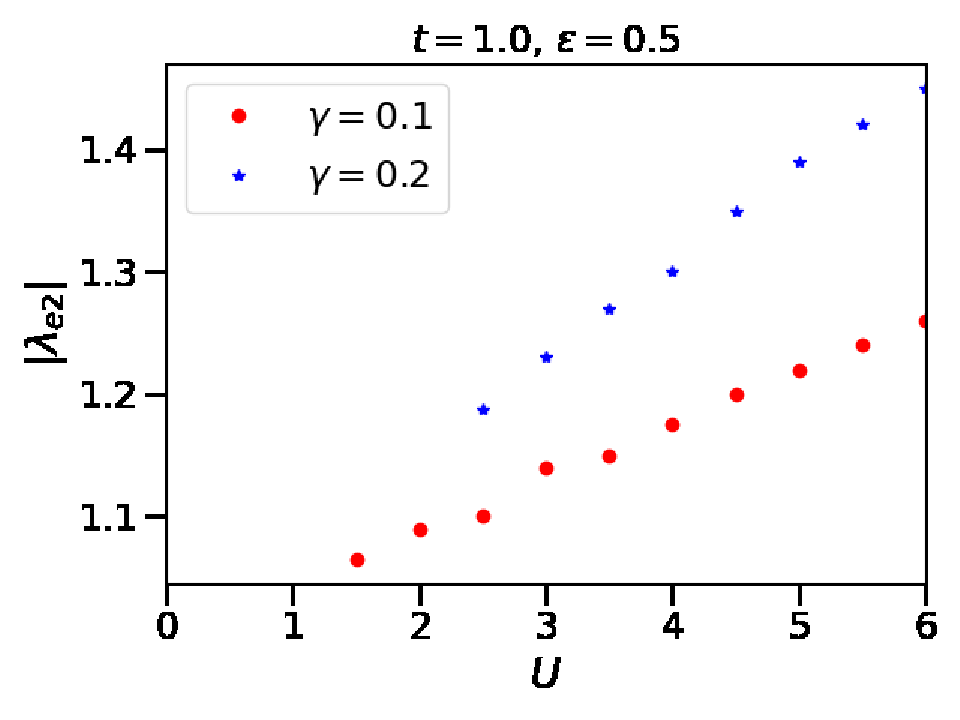}
		 \caption{}
		 \label{fig:lambdae2:vs:U}
	\end{subfigure}%
	\begin{subfigure}[t]{.5\linewidth}
		\centering
		\includegraphics[height=\myht]{\FIGDIR/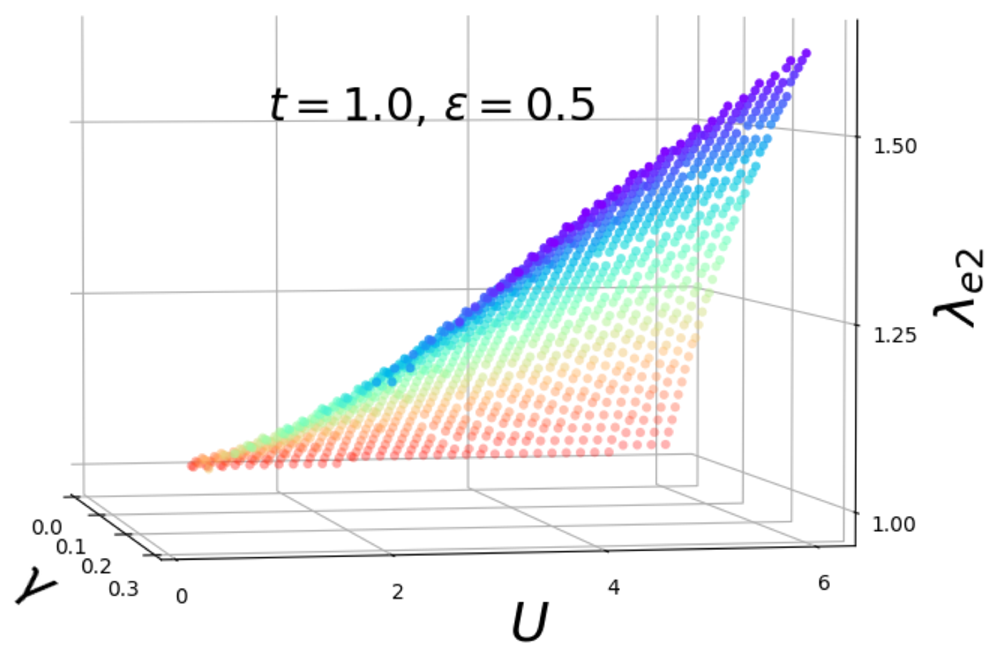}%
		 \caption{}
		 \label{fig:lambdae2:vs:gamma:U}
	\end{subfigure}%
        \hfill
        \begin{subfigure}[t]{.5\linewidth}
		\centering
		\includegraphics[height=\myht]{\FIGDIR/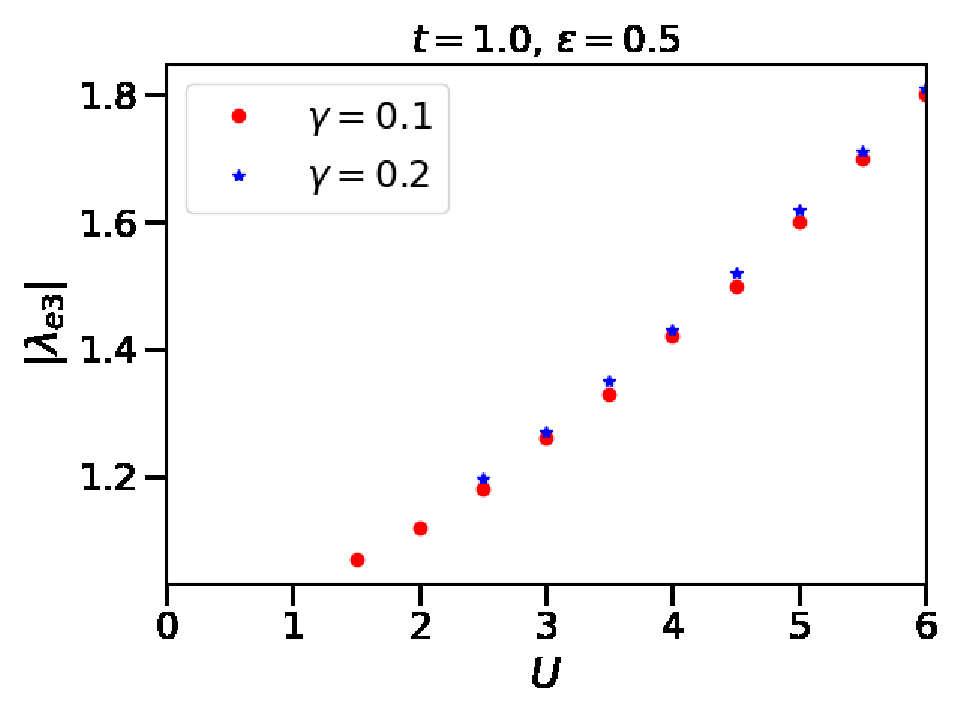}
		 \caption{}
		 \label{fig:lambdae3:vs:U}
	\end{subfigure}%
	\begin{subfigure}[t]{.5\linewidth}
		\centering
		\includegraphics[height=\myht]{\FIGDIR/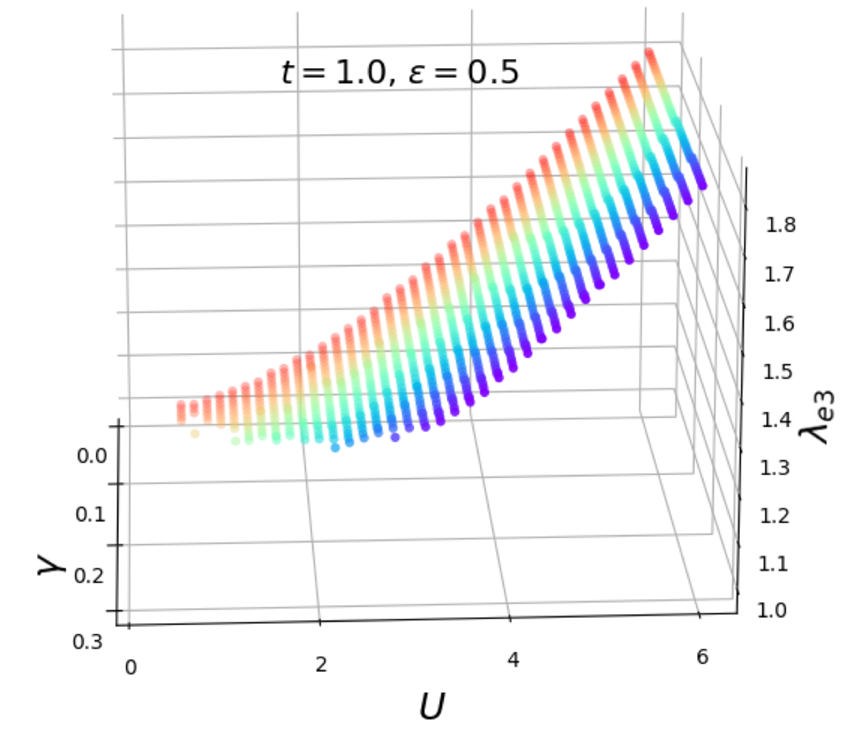}%
		 \caption{}
		 \label{fig:lambdae3:vs:gamma:U}
	\end{subfigure}%
	\caption{Positions of exceptional points (a) $\lambda_{e1}$, (c) $\lambda_{e2}$, and (e) $\lambda_{e2}$ as Hubbard interaction $U$ is varied for different $\gamma$'s at $t=1$, $\eps=0.5$. Three dimensional plots of them over $\g$-$U$ plane are shown in (b), (d), and (f) respectively.}
\label{fig:lambdae1:lambdae2:lambdae3:vs:U}
\end{figure}

\section{Physical interpretation and outlook}
$\PT$ symmetric non-Hermitian physics have been successfully observed in several two level photonic and optical systems~\cite{book:bender18:ptsymmetry}. 
One particular feature of such Hamiltonians is the existence of exceptional points (EPs) beyond which complex eigenenergies emerge signaling breaking of the symmetry in the eigenfunctions. As a simplistic model, we consider a hydrogen molecule with Hubbard interaction acting between its atoms' electrons. 
\col{We then introduce both diagonal and off-diagonal non-Hermitian terms to the Hamiltonian preserving its $\PT$ symmetry.
Though effects of diagonal and off-diagonal terms have been studied extensively in the past, however, discussions on their dual presence in the same Hamiltonian is surprisingly missing in the literature, specifically with a particle-particle interaction term. Our work fills up this fundamental gap.}
 
\col{
In our model, the $\PT$-symmetry broken phases occur due to competitions among several
energy scales, mainly $t$, $U$, $\lambda$, and $\gamma$ (in our treatment, the orbital energy $\eps$ is a reference scale which can be set to zero or any constant value without loss of generality). $t$ and $U$ constitute the Hermitian terms, however, they compete against each other and take off-diagonal and diagonal positions respectively in the Hamiltonian matrix. Like in any $\PT$-symmetric Hamiltonian, both non-Hermiticity parameters $\lambda$ and $\gamma$ compete with the Hermitian $t$-term and spontaneously  break the $\PT$-symmetry when the eigenenergies become complex (terms inside the square root becomes negative). The eigenenergy plots for both parameters are similar and follow the generic two-level state (TLS)'s $\PT$-symmetric diagrams. However, the presence of Hubbard interaction $U$ destroys the similarity between $\lambda$ and $\gamma$ and gives rise to a complicated cubic equation, which generically can have two complex roots in the broken symmetry regime. 
It is the nature of the Hubbard interaction and fermionic operators that gives rise to the $6\times 6$ matrix (\eref{eq:H3:Sz:0:matrix}) which reduces to a cubic characteristic equation (for $U\ne 0$) instead of a typical TLS-type quadratic one.}
\col{
The physical understanding of this scenario could be the following. 
In the absence of interaction and off-diagonal non-Hermiticity $\lambda$, $\gamma$-term has to compete with the symmetric hopping $t$ only. Hence $U=0$ places the $\PT$-breaking boundary at the hopping amplitude: $|\gamma_e|=t$ (\fref{fig:gamma_e:vs_U}). In the TLS language, $t$ induces \emph{avoided level crossing} while $\gamma$ encourages crossing or transition between two energy levels~\cite{wrona:etal:srep20}.
Since Hubbard interaction $U$ goes against the kinetic term $t$, the term eventually enhances the effect of $\gamma$ (dissociation of the $H_2$ molecule)~\cite{wrona:etal:srep20}. Thus we see  $|\gamma_e|$ monotonically decreasing as $U$'s value rises. Now finite $\lambda$ joins hands with $\gamma$ and together lead to dynamical instability, hence $PT$-boundary gets reduced in the presence of finite $\lambda$ (\fref{fig:gamma_e:vs_U}). It is worth mentioning that recently Pan and coworkers~\cite{pan:wang:cui:chen:pra20}] studied the Hubbard dimer model using the Lindblad master equation and found that the rescaled probability diverges at the same threshold value of interaction strength as found in the $\PT$-breaking effective non-Hermitian Hubbard dimer Hamiltonian. 
}   

We notice that interaction plays differently with different kinds of EPs generated by the parameters of the Hamiltonian. Changing the position of one kind of EPs in the increasing direction and the other kind in decreasing direction by varying interaction strength can offer flexibility in fine tuning EPs and more control over their potential applications. In a realistic hydrogen molecule, non-Hermitian loss-gain terms might be introduced through laser induced molecular ionization and dissociation~\cite{lefebvre:etal:prl09,wrona:etal:srep20}. Besides this, a more precise two-site Hubbard model could be emulated in an ultracold double well system~\cite{murmann:etal:prl15} or via  NMR~\cite{melo:etal:nmr:po21}.
The role of Hubbard interaction on the EPs has been studied recently~\cite{pan:wang:cui:chen:pra20}. However, the interplay of the diagonal and off-diagonal $\PT$-symmetries and the role
of interaction on them have not been studied ever to the best of our knowledge. Such interplay might be extended to the fermionic or bosonic lattice Hubbard models and effect on interesting physics such as closure of Mott gap~\cite{tripathi:galda:barman:vinokur:prb16,tripathi:vinokur:srep20,rausch:etal:njp21} or multiple $\PT$-broken phases~\cite{jin:song:ap13} can be studied. Finally, we must mention that the cube-root based EPs, found in our simplistic model, may require deeper understanding with potential applications in various areas of physics and quantum chemistry.

\section{Acknowledgement and announcement}
The authors thank the HBCSE, Mumbai, for providing an opportunity to collaborate through their NIUS Physics 15.3 camp. HB is indebted to Vikram Tripathi, TIFR, Mumbai, for several useful discussions and his valuable inputs.

Our codes are available on the Github repository: \newline\url{https://github.com/hbaromega/PT-symmetric-2-site-Hubbard-hydrogen}, \newline under GNU General Public License. 

%
\appendix
\section{Construction of non-zero matrix elements of $H^0$}
\label{app:construct:H0}
For a 2-site electronic system, $4^2=16$ possible atomic states can appear which can be denoted as
\\ 
\\
$\ket{1}\equiv\ket{0,0}$, $\ket{2}\equiv\ket{\ua,0}$, $\ket{3}\equiv\ket{\da,0}$, $\ket{4}\equiv\ket{\ua\da,0}$,\\
$\ket{5}\equiv\ket{0,\ua}$, $\ket{6}\equiv\ket{\ua,\ua}$, $\ket{7}\equiv\ket{\da,\ua}$, $\ket{8}\equiv\ket{\ua\da,\ua}$,\\
$\ket{9}\equiv\ket{0,\da}$, $\ket{10}\equiv\ket{\ua,\da}$, $\ket{11}\equiv\ket{\da,\da}$, $\ket{12}\equiv\ket{\ua\da,\da}$,\\
$\ket{13}\equiv\ket{0,\ua\da}$, $\ket{14}\equiv\ket{\ua,\ua\da}$, $\ket{15}\equiv\ket{\da,\ua\da}$, $\ket{16}\equiv\ket{\ua\da,\ua\da}$\\
\\
while in the state $\ket{\alpha,\beta}$, $\alpha$ and $\beta$ represent the states of 
site (or atom) 1 and 2 respectively. Also, we stick to a convention that when two fermionic operators 
operate together, the site-1 operator acts first, i.e. it has to always be brought to 
the right of the site-2 operator. For example,
\blgn
c\y_{2\da}c\y_{1\ua}\ket{0,0}=\ket{\ua,\da}\,
\label{eq:example:1}
\elgn
This distinguishes from the other possible action of the same operators together but in the reverse
order (by a minus factor):
\blgn
c\y_{1\da}c\y_{2\ua}\ket{0,0}=-\ket{\ua,\da}\,
\label{eq:example:2}
\elgn
respecting the fermionic anticommutation rule
\blgn
\{c\y_{1\alpha},c\y_{2\beta}\}=c\y_{1\alpha}c\y_{2\beta} + c\y_{2\beta}c\y_{1\alpha}=0\,.
\elgn

{\it Convention: site-1 operator acts first and among two same site operators 
of different spins, $\ua$-spin operator
will be prior to act.}

For a hydrogen molecule, total number of electrons is $N=2$. Therefore to form the basis, we need to only consider ${}^4C_2=6$ states restricted to $N=2$:
\blgn
\ket{1}&\equiv \ket{\ua,\ua}=c\y_{2\ua}c\y_{1\ua}\ket{0}\\
\ket{2}&\equiv \ket{0,\ua\da}=c\y_{2\da}c\y_{2\ua}\ket{0}\\
\ket{3}&\equiv \ket{\ua,\da}=c\y_{2\da}c\y_{1\ua}\ket{0}\\
\ket{4}&\equiv \ket{\da,\ua}=c\y_{2\ua}c\y_{1\da}\ket{0}\\
\ket{5}&\equiv \ket{\ua\da,0}=c\y_{1\da}c\y_{1\ua}\ket{0}\\
\ket{6}&\equiv \ket{\da,\da}=c\y_{2\da}c\y_{1\da}\ket{0}\,.
\elgn
To construct the Hamiltonian in matrix form, we operate the Hamiltonian $\h H$
on each of the 6 states and we find 
\be
\boxed{
\h H^0 \ket{1}=2\eps\, c\y_{2\ua}c\y_{1\ua}\ket{0}=2\eps\ket{1}\,.
}
\ee
\blgn
\h H^0 \ket{2}&=2\eps\, c\y_{2\ua}c\y_{2\ua}\ket{0}+t(c\y_{1\da}c\py_{2\da}+c\y_{1\ua}c\py_{2\ua})
c\y_{2\da}c\y_{2\ua}\ket{0}\non\,.
\elgn
%
%
%
%
%
%
We notice 
\blgn
c\y_{1\da}c\py_{2\da} c\y_{2\da}c\y_{2\ua}\ket{0}
&=c\y_{1\da} (1-c\y_{2\da}c\py_{2\da}) c\y_{2\ua}\ket{0}\non\\
&\qquad\mbox{[Used $\{c\py_{2\da},c\y_{2\da}\}=1$]}\non\\
&=c\y_{1\da} c\y_{2\ua}\ket{0}\non\\
&=-c\y_{2\ua}c\y_{1\da}\ket{0}\non\\
&=-\ket{4}\non
\elgn

and 
\blgn
c\y_{1\ua}c\py_{2\ua} c\y_{2\da}c\y_{2\ua}\ket{0}
&=-c\y_{1\ua}c\py_{2\ua} c\y_{2\ua}c\y_{2\da}\ket{0}\non\\
&=-c\y_{1\ua} (1-c\y_{2\ua} c\py_{2\ua}) c\y_{2\da}\ket{0}\non\\
&\qquad\mbox{[Used $\{c\py_{2\ua},c\y_{2\ua}\}=1$]}\non\\
&=-c\y_{1\ua} c\y_{2\da}\ket{0}\non\\
&=c\y_{2\da}c\y_{1\ua} \ket{0}\non\\
&=\ket{3}\non\,.
\elgn

Thus
\be
\boxed{
\h H^0 \ket{2}=2\eps\ket{2}+t\,(\ket{3}-\ket{4})\,.
}
\ee

By proceeding in the same fashion, we find



%
\be
\boxed{
\h H^0 \ket{3}=2\eps\ket{3}+t\,(\ket{2}+\ket{5})\,.
}
\ee
%



\be
\boxed{
\h H^0 \ket{4}=2\eps\ket{4}+t\,(-\ket{2}-\ket{5})\,.
}
\ee
%



\be
\boxed{
\h H^0 \ket{5}=2\eps\ket{5}+t\,(\ket{3}-\ket{4})\,.
}
\ee
\be
\boxed{
\h H^0 \ket{6}=2\eps\,c\y_{2\da}c\y_{1\da}\ket{0}=2\eps\ket{6}\,.
}
\ee


Hence the Hamiltonian in matrix form: 
\begin{align}
{\bf H^0}=
  \begin{bmatrix}
     2\eps &0     &0      &0     &0     &0 \\  
     0     &2\eps &t      &-t    &0     &0 \\  
     0     &t     &2\eps  &0     &t     &0 \\  
     0     &-t    &0      &2\eps &-t    &0 \\  
     0     &0     &t      &-t     &2\eps &0 \\  
     0     &0     &0      &0     &0     &2\eps
  \end{bmatrix}
  \,.
\end{align}

\section{Construction of non-zero matrix elements of $H^1$:} 
\label{app:construct:H1}
We use the same basis as before, and repeat the steps followed in ~\ref{app:construct:H0}.
Thus
\blgn
\h H^1 \ket{1} &=2\eps c\y_{2\ua}c\y_{1\ua}\ket{0}=2\eps\ket{1}\,.\\
\h H^1 \ket{2} &=2\eps\ket{2}+t^+\,(\ket{3}-\ket{4})\,.\\
\h H^1 \ket{3} &=2\eps\ket{3}+t^-\ket{2}+t^+\ket{5}\,.\\
\h H^1 \ket{4} &=2\eps\ket{4}-t^-\ket{2}-t^+\ket{5}\,.\\
\h H^1 \ket{5} &=2\eps\ket{5}+t^-\,(\ket{3}-\ket{4})\,.\\
\h H^1 \ket{6} &=2\eps c\y_{2\da}c\y_{1\da}\ket{0}=2\eps\ket{6}\,.
\elgn

Therefore the nonzero matrix elements in the Hamiltonian are
\blgn
H^1_{11}&=H'_{22}=H'_{33}=H'_{44}=H'_{55}=H'_{66}=2\eps\\
H^1_{23}&=\tm\\
H^1_{32}&=\tp\\
H^1_{24}&=-\tm\\
H^1_{42}&=-\tp\\
H^1_{35}&=\tm\\
H^1_{53}&=\tp\\
H^1_{45}&=-\tm\\
H^1_{54}&=-\tp\\
\elgn
Hence the Hamiltonian in matrix form: 
\begin{align}
{\bf H^1}=
  \begin{bmatrix}
     2\eps &0     &0      &0     &0     &0 \\  
     0     &2\eps &t^-      &-t^-    &0     &0 \\  
     0     &t^+     &2\eps  &0     &t^-     &0 \\  
     0     &-t^+    &0      &2\eps &-t^-    &0 \\  
     0     &0     &t^+      &-t^+     &2\eps &0 \\  
     0     &0     &0      &0     &0     &2\eps
  \end{bmatrix}
  \,.
\label{eq:H1:Sz:0:matrix}
\end{align}

\section{Simplification of the cubic equation}
\label{app:cubic}

\eref{eq:cubic:doub:diss:Hubb} of the main text can be simplified as
\blgn
&(\esum-\ediff+U-E)(\esum-E)(\esum+\ediff+U-E)\non\\
&\quad-M(\esum-E)-MU=0\non\\
&\quad
\mbox{[Define: $\esum\equiv \epsp+\epsm$, $\ediff \equiv \epsp-\epsm$, $M\equiv 4\tp\tm$]}\non\\
&\Ra (x-D+U)x(x+D+U)-Mx-MU=0\,.\non\\
&\qquad\mbox{[Define: $x\equiv S-E$]}\non\\
&\Ra [(x+U)^2-D^2]x-M(x+U)=0\,.\non\\
&\Ra [X^2-D^2](X-U)-MX=0\,.\non\\
&\qquad\mbox{[Define: $X\equiv x+U$]}\non\\
&\Ra X^3-U X^2-(D^2+M)X+D^2 U=0\,,\non\\
&\Ra X^3-U X^2-K X- L=0\,\tag{\ref{eq:doub:diss:Hubb:cubic:form}}\\
&\qquad\mbox{[Define: $K\equiv D^2+M$; $L\equiv -D^2U$]}\,.\non
\elgn

\section{Convention of marking energy bands}
\label{app:marking:convention}
\subsection*{Noninteracting case:}
\begin{figure}[hbp!]
\begin{subfigure}[t]{.6\linewidth}
\centering\includegraphics[height=6cm,clip]{\FIGDIR/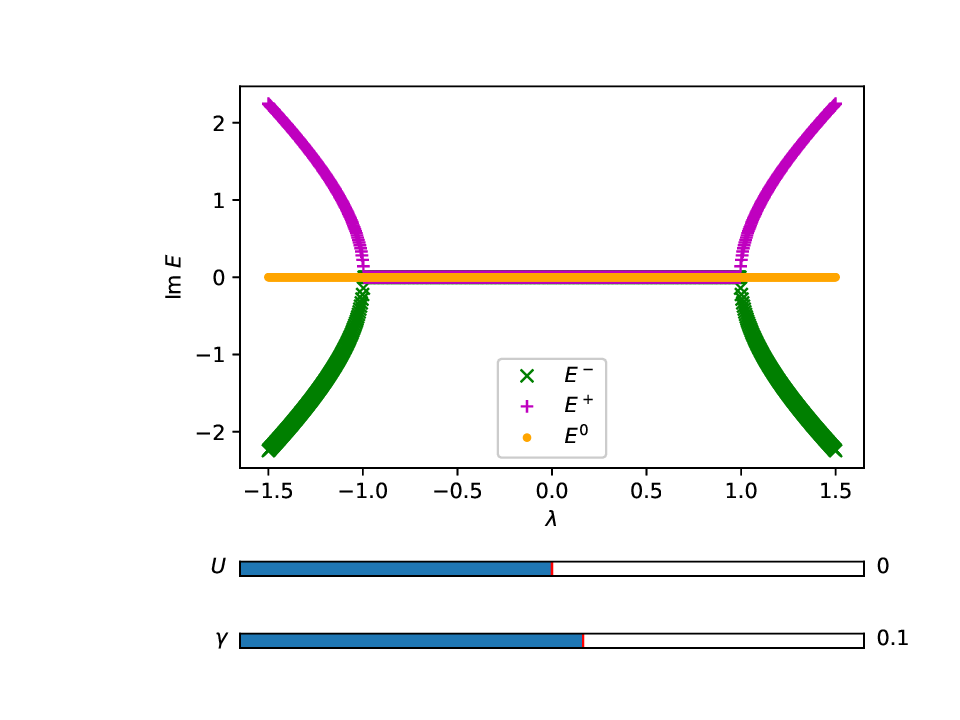}
\caption{}
\label{fig:ImE:vs:lambda:U0:g0.1}
\end{subfigure}
\begin{subfigure}[t]{.6\linewidth}
\centering\includegraphics[height=6cm,clip]{\FIGDIR/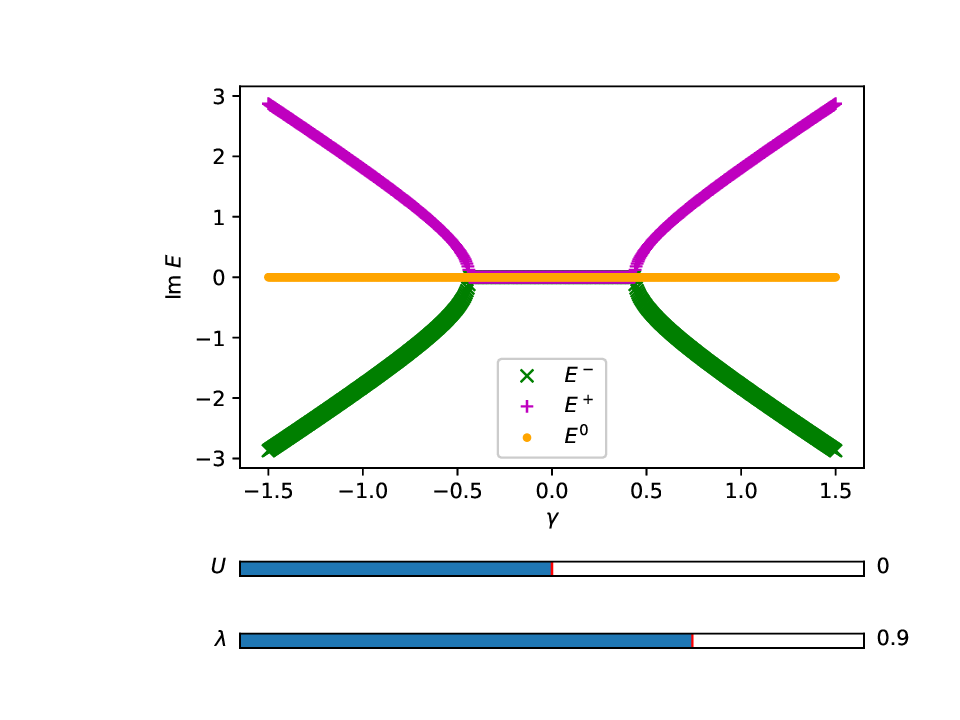}
\caption{}
\label{fig:ImE:vs:gamma:U0:l0.9}
\end{subfigure}
\caption{Imaginary parts of the three eigenergies plotted against (a) $\lambda$ for fixed $\gamma$ and (b)
 $\gamma$ for fixed $\lambda=0.9$ for the non-interacting case ($U=0$).}
\end{figure}

The key equation (characteristic equation) whose roots provide the eigenenergies:
\blgn
X^3-U X^2-K X - L=0
\tag{\ref{eq:doub:diss:Hubb:cubic:form}}
\elgn
with $X\equiv x+U$; $x\equiv 2\eps-E$; $K\equiv 4(t^2-\g^2-\l^2)$; $L\equiv 4\g^2 U$. 
When $U=0$, \eref{eq:doub:diss:Hubb:cubic:form} effectively becomes quadratic as $L=4\g^2 U$ becomes zero. However, keeping the original cubic equation in mind, we can state that \eref{eq:doub:diss:Hubb:cubic:form} has one trivial root $X=x=0$ or eigenenergy $E=2\eps=1$ for $\eps=0.5$ and $\forall\,\l,\g$. 
Thus for $U=0$, we end up with two semicircular bands and a 
third one which stays always constant for the real parts of the eigenenergies ($\re\,E=1$). We assign {\red red} color for the upper semicircle ($E^+$), {\blue blue} ($E^-$) for the lower one and {\cyan cyan} for the constant line ($E^0$) 
%
%
\begin{figure}[hbp!]
\begin{subfigure}[t]{.6\linewidth}
\centering\includegraphics[height=6cm,clip]{\FIGDIR/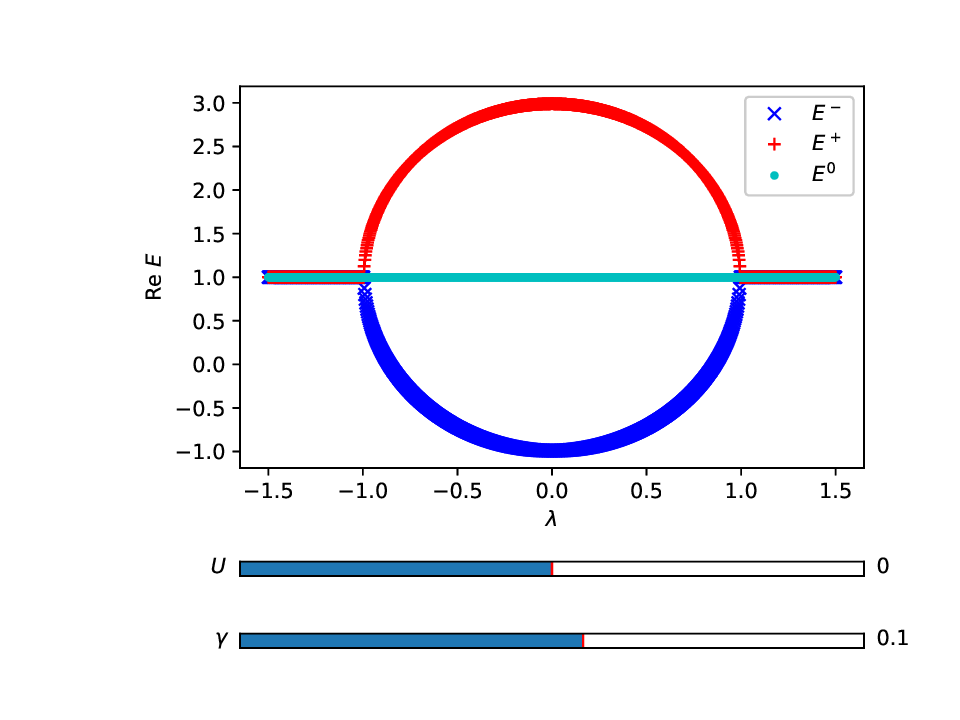}
\caption{}
\label{fig:ReE:vs:lambda:U0:g0.1}
\end{subfigure}
\begin{subfigure}[t]{.6\linewidth}
\centering\includegraphics[height=6cm,clip]{\FIGDIR/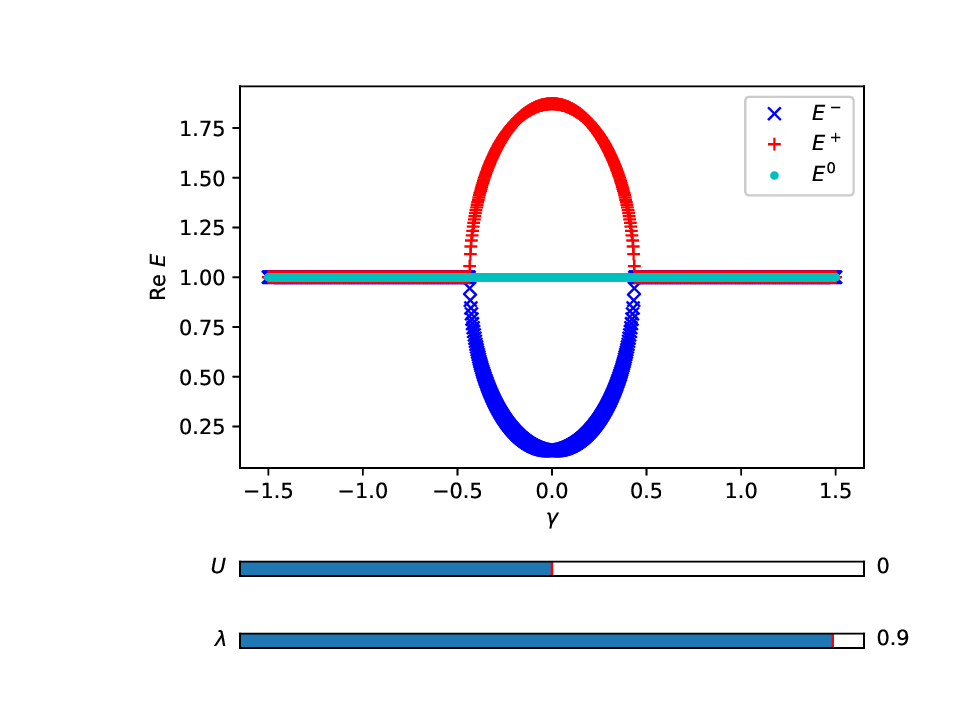}
\caption{}
\label{fig:ReE:vs:gamma:U0:l0.9}
\end{subfigure}
\caption{Real parts of the three eigenergies plotted against (a) $\lambda$ for fixed $\gamma$ and (b)
 $\gamma$ for fixed $\lambda=0.9$ for the non-interacting case ($U=0$).}
\end{figure}
(see \fref{fig:ReE:vs:lambda:U0:g0.1} and \fref{fig:ReE:vs:gamma:U0:l0.9}).
To identify the bands for the offline (black and white) version, we use symbols: `plus'($+$), `cross'($\times$), and `filled circle/dot'($\bullet$) respectively.
Similarly, for the imaginary parts of the eigenenergies, we color the bands with positive value by {\magenta magenta}, negative value by {\dgreen green}, and zero by {\orange orange} (see \fref{fig:ImE:vs:lambda:U0:g0.1} and \fref{fig:ImE:vs:gamma:U0:l0.9}). 
%
%
We make the following correspondence between the colors of the real and imaginary parts of the eigenenergies.

\begin{center}
\begin{tabular}{c|c|c}
 $\re E$ $\to$   $\im E$  & Label &Symbol\\
 \hline
 {\red red} $\to$   {\magenta magenta} &$E^+$ &$+$\\
 {\blue blue} $\to$  {\dgreen green}  &$E^-$ &$\times$\\
 {\cyan cyan} $\to$  {\orange orange} &$E^0$ &$\bullet$\\
 \hline
\end{tabular}
\end{center}

This mapping needs to be obeyed to see how an eigenenergy band as a complex entity change 
as a function of the parameter $\lambda$ or $\gamma$. Despite having this color convention, as we turn on $U$, keeping track of the eigenenergies (as continuous functions of the parameters) becomes tedious and we encounter ambiguities when overlaps occur in the real or imaginary parts of two eigenenergies as described below.

\subsection*{Finite interaction case:}
\begin{figure}[hbp!]
\begin{subfigure}[t]{.6\linewidth}
\includegraphics[height=5.5cm,clip]{\FIGDIR/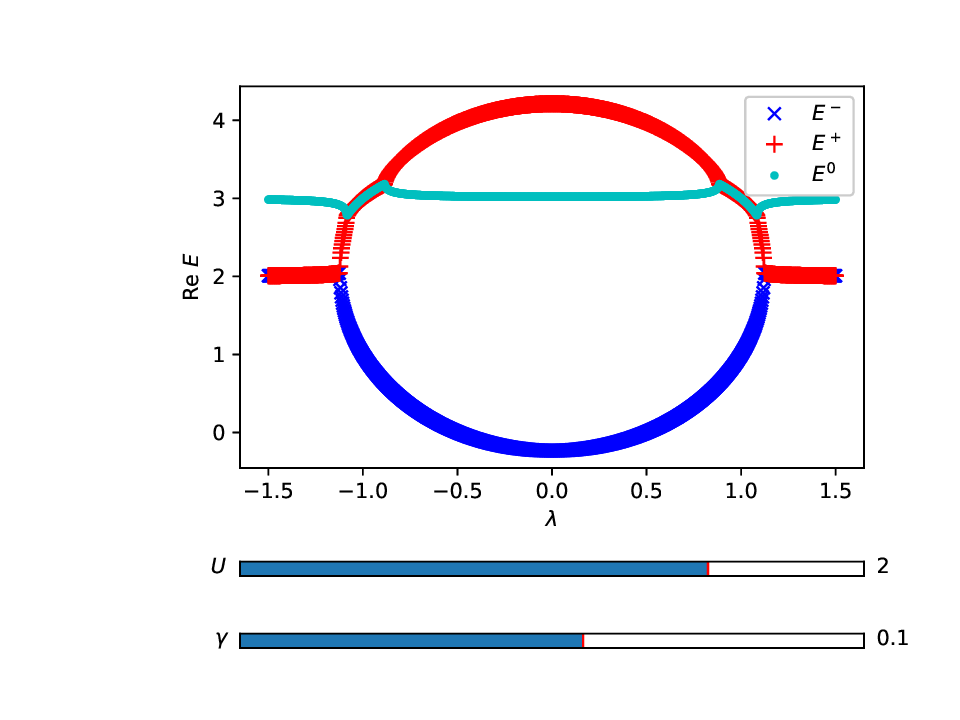}
\caption{}
\label{fig:ReE:vs:lambda:U2:g0.1}
\end{subfigure}
\begin{subfigure}[t]{.6\linewidth}
\includegraphics[height=5.5cm,clip]{\FIGDIR/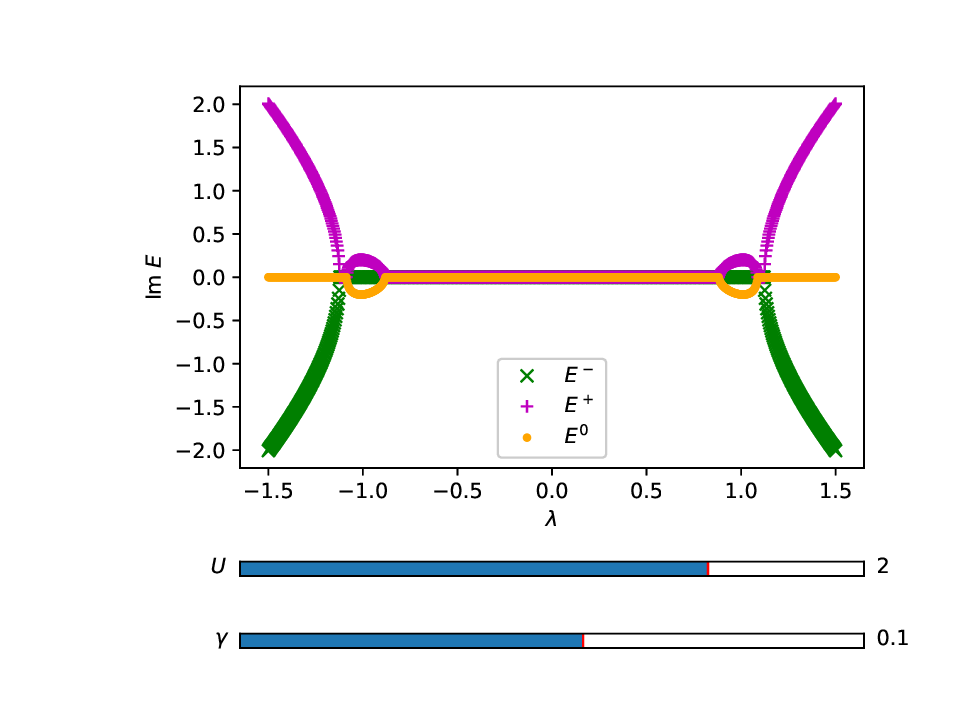}
\caption{}
\label{fig:ImE:vs:lambda:U2:g0.1}
\end{subfigure}
\caption{(a) Real and (b) imaginary parts of the three eigenergies plotted against $\lambda$ for fixed $\gamma = 0.1$ for finite Hubbard interaction ($U=2$).}
\end{figure}

We take the example of plotting eigenenergies against $\lambda$ for $U=2$ and $\gamma=0.1$. Here we
encounter the cases where ambiguity in band-coloring arises due to the breaking of mirror symmetry 
between $E^+$ and $E^-$ at a finite value of $U$. Below we mention them and discuss how we 
resolve the ambiguity by selecting a reasonable convention after taking a cue from that in the $U=0$ case. 
Let us consider the following cases in the eigenenergy plots for $U=2$ and $\g=0.1$ (see \fref{fig:ReE:vs:lambda:U2:g0.1} and \fref{fig:ImE:vs:lambda:U2:g0.1}).

%
%

{\bf Case 1:}  Two real values of the eigenenergies (colored red and blue) overlap ($|\lambda| > 1$), 
 positive and negative values of $\im E$
can be painted either magenta  or green. 
Let us choose magenta ($\Ep$) for the positive values of $\im E$ and green ($\Em$) for the negative one. 
 
{\bf Case 2:} Two merged real eigenenergies branching out into two separate eigenvalues (purely real, imaginary parts for both are zero, $|\lambda| < 1$).
 Of course, the colors of the two branches could be either red or blue. Following the $U=0$ convention, 
we paint the upper band with red ($\Ep$) and the lower one with blue ($\Em$). 

{\bf Case 3:} Again two real eigenenergies merge or overlap each other (this time, red and cyan, 0.88 < $|\lambda| < 1.05$). So ambiguity arises in coloring the pair of complex conjugates of $\im E$. We stick to the convention of painting magenta ($\Ep$) for the curve having positive values. So the lower negative curve ($\Ez$) becomes orange in color. 

{\bf Case 4:} Again branching happens in the real parts as the imaginary parts become around $|\lambda| < 0.88$.
 The colors of the branches become ambiguous. Sticking to the convention, we paint the upper band with red ($\Ep$) and the lower one with cyan ($\Ez$). 

\subsection*{Shift of $E^0$ band due to finite interaction:}
\begin{figure}[htp!]
\begin{subfigure}[t]{.6\linewidth}
\includegraphics[height=5cm,clip]{\FIGDIR/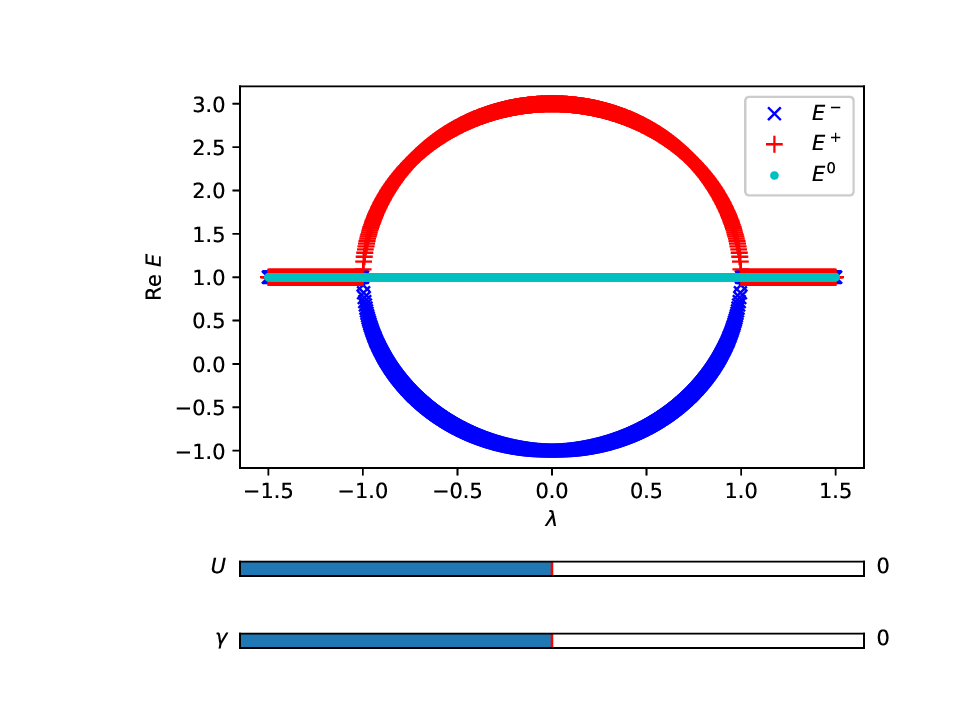}
\caption{}
\label{fig:ReE:vs:lambda:U0:g0}
\end{subfigure}
\begin{subfigure}[t]{.6\linewidth}
\includegraphics[height=5cm,clip]{\FIGDIR/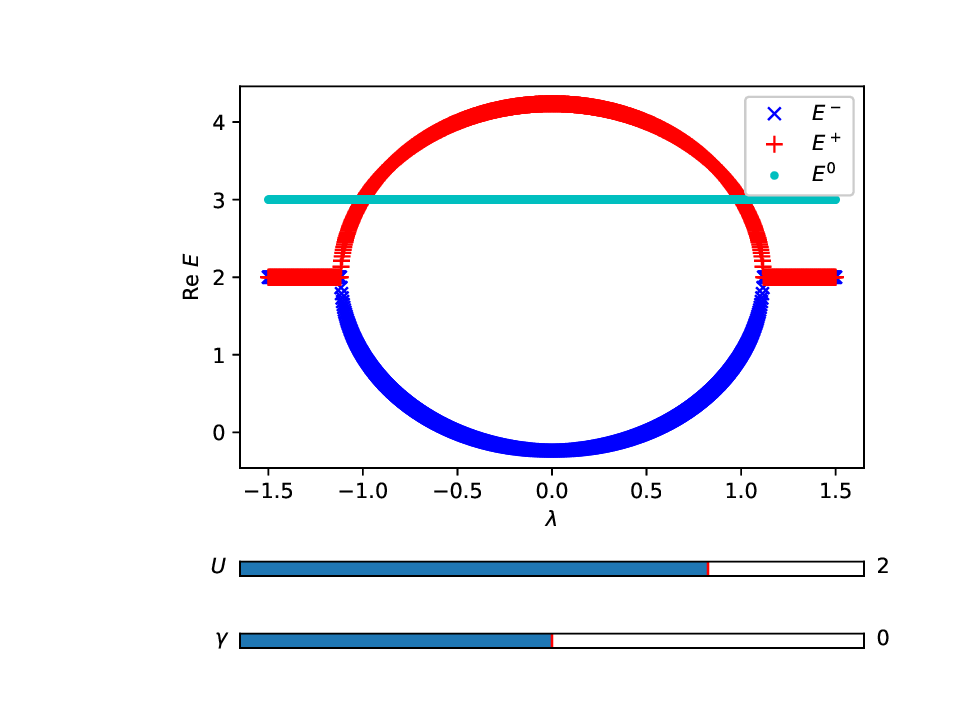}
\caption{}
\label{fig:ReE:vs:lambda:U2:g0}
\end{subfigure}
\caption{Real parts of the three eigenergies plotted against $\lambda$ for $\gamma=0$ for two cases: (a) $U=0$ and (b) $U=2$.}
\end{figure}

We know that \eref{eq:doub:diss:Hubb:cubic:form} also reduces to a quadratic equation as $L$ becomes zero again. Thus both presence of interaction and diagonal non-Hermiticity are essential to see the second kind of EPs. By turning off $\gamma$, we can easily see that interaction only shifts the position of $E^0$ (by an amount $U$, see \fref{fig:ReE:vs:lambda:U0:g0} and \fref{fig:ReE:vs:lambda:U0:g0}). Thus real part must always lie between real parts of $E^+$ and $E^-$ bands (in the extreme case, it may overlap with $E^+$ for $U>0$). Presence of finite $\g$ creates an overlap between $E^+$ and $E^0$ for $U>0$ and we obtain additional EPs. This observation provides the confidence to keep the $\re\,E^0$ (cyan) curve in between $\re\,E^+$ (red) and $\re\,E^-$ (blue).
%
%

\section*{References}
\bibliographystyle{unsrt}
\bibliography{refs_nonHerm_H2}

\end{document}